\documentclass[superscriptaddress,prl,twocolumn]{revtex4-1}
\pdfoutput=1
\usepackage{amsmath,mathrsfs,amsbsy,color,graphicx,bm,amsthm,amsfonts,dsfont}
\usepackage{times}
\usepackage{units}
\usepackage{braket}
\usepackage{epstopdf}
\usepackage{amscd}
\usepackage{enumerate}
\usepackage{multirow}

\begin{document}
\title{Observation of Ten-photon Entanglement Using Thin BiB$_{3}$O$_{6}$ Crystals}

\author{Luo-Kan Chen}
\affiliation{Shanghai Branch, National Laboratory for Physical Sciences at Microscale and Department of Modern Physics, University of Science and Technology of China, Hefei, Anhui 230026, China}
\affiliation{CAS Center for Excellence and Synergetic Innovation Center in Quantum Information and Quantum Physics,
University of Science and Technology of China, Shanghai 201315, China}
\affiliation {CAS-Alibaba Quantum Computing Laboratory, Shanghai, 201315, China}

\author{Zheng-Da Li}
\affiliation{Shanghai Branch, National Laboratory for Physical Sciences at Microscale and Department of Modern Physics, University of Science and Technology of China, Hefei, Anhui 230026, China}
\affiliation{CAS Center for Excellence and Synergetic Innovation Center in Quantum Information and Quantum Physics,
University of Science and Technology of China, Shanghai 201315, China}
\affiliation {CAS-Alibaba Quantum Computing Laboratory, Shanghai, 201315, China}

\author{Xing-Can Yao}
\affiliation{Shanghai Branch, National Laboratory for Physical Sciences at Microscale and Department of Modern Physics, University of Science and Technology of China, Hefei, Anhui 230026, China}
\affiliation{CAS Center for Excellence and Synergetic Innovation Center in Quantum Information and Quantum Physics,
University of Science and Technology of China, Shanghai 201315, China}
\affiliation {CAS-Alibaba Quantum Computing Laboratory, Shanghai, 201315, China}

\author{Miao Huang}
\affiliation{Shanghai Branch, National Laboratory for Physical Sciences at Microscale and Department of Modern Physics, University of Science and Technology of China, Hefei, Anhui 230026, China}
\affiliation{CAS Center for Excellence and Synergetic Innovation Center in Quantum Information and Quantum Physics,
University of Science and Technology of China, Shanghai 201315, China}
\affiliation {CAS-Alibaba Quantum Computing Laboratory, Shanghai, 201315, China}

\author{Wei Li}
\affiliation{Shanghai Branch, National Laboratory for Physical Sciences at Microscale and Department of Modern Physics, University of Science and Technology of China, Hefei, Anhui 230026, China}
\affiliation{CAS Center for Excellence and Synergetic Innovation Center in Quantum Information and Quantum Physics,
University of Science and Technology of China, Shanghai 201315, China}
\affiliation {CAS-Alibaba Quantum Computing Laboratory, Shanghai, 201315, China}

\author{He Lu}
\affiliation{Shanghai Branch, National Laboratory for Physical Sciences at Microscale and Department of Modern Physics, University of Science and Technology of China, Hefei, Anhui 230026, China}
\affiliation{CAS Center for Excellence and Synergetic Innovation Center in Quantum Information and Quantum Physics,
University of Science and Technology of China, Shanghai 201315, China}
\affiliation {CAS-Alibaba Quantum Computing Laboratory, Shanghai, 201315, China}

\author{Xiao Yuan}
\affiliation{Center for Quantum Information, Institute for Interdisciplinary Information Sciences, Tsinghua University, Beijing 100084, China}

\author{Yan-Bao Zhang}
\affiliation{Institute for Quantum Computing and Department of Physics and Astronomy, University of Waterloo, Waterloo, Ontario, N2L 3G1 Canada}

\author{Xiao Jiang}
\affiliation{Shanghai Branch, National Laboratory for Physical Sciences at Microscale and Department of Modern Physics, University of Science and Technology of China, Hefei, Anhui 230026, China}
\affiliation{CAS Center for Excellence and Synergetic Innovation Center in Quantum Information and Quantum Physics,
University of Science and Technology of China, Shanghai 201315, China}
\affiliation {CAS-Alibaba Quantum Computing Laboratory, Shanghai, 201315, China}

\author{Cheng-Zhi Peng}
\affiliation{Shanghai Branch, National Laboratory for Physical Sciences at Microscale and Department of Modern Physics, University of Science and Technology of China, Hefei, Anhui 230026, China}
\affiliation{CAS Center for Excellence and Synergetic Innovation Center in Quantum Information and Quantum Physics,
University of Science and Technology of China, Shanghai 201315, China}
\affiliation {CAS-Alibaba Quantum Computing Laboratory, Shanghai, 201315, China}

\author{Li Li}
\affiliation{Shanghai Branch, National Laboratory for Physical Sciences at Microscale and Department of Modern Physics, University of Science and Technology of China, Hefei, Anhui 230026, China}
\affiliation{CAS Center for Excellence and Synergetic Innovation Center in Quantum Information and Quantum Physics,
University of Science and Technology of China, Shanghai 201315, China}
\affiliation {CAS-Alibaba Quantum Computing Laboratory, Shanghai, 201315, China}

\author{Nai-Le Liu}
\affiliation{Shanghai Branch, National Laboratory for Physical Sciences at Microscale and Department of Modern Physics, University of Science and Technology of China, Hefei, Anhui 230026, China}
\affiliation{CAS Center for Excellence and Synergetic Innovation Center in Quantum Information and Quantum Physics,
University of Science and Technology of China, Shanghai 201315, China}
\affiliation {CAS-Alibaba Quantum Computing Laboratory, Shanghai, 201315, China}

\author{Xiongfeng Ma}
\affiliation{Center for Quantum Information, Institute for Interdisciplinary Information Sciences, Tsinghua University, Beijing 100084, China}

\author{Chao-Yang Lu}
\affiliation{Shanghai Branch, National Laboratory for Physical Sciences at Microscale and Department of Modern Physics, University of Science and Technology of China, Hefei, Anhui 230026, China}
\affiliation{CAS Center for Excellence and Synergetic Innovation Center in Quantum Information and Quantum Physics,
University of Science and Technology of China, Shanghai 201315, China}
\affiliation {CAS-Alibaba Quantum Computing Laboratory, Shanghai, 201315, China}

\author{Yu-Ao Chen}
\affiliation{Shanghai Branch, National Laboratory for Physical Sciences at Microscale and Department of Modern Physics, University of Science and Technology of China, Hefei, Anhui 230026, China}
\affiliation{CAS Center for Excellence and Synergetic Innovation Center in Quantum Information and Quantum Physics,
University of Science and Technology of China, Shanghai 201315, China}
\affiliation {CAS-Alibaba Quantum Computing Laboratory, Shanghai, 201315, China}

\author{Jian-Wei Pan}
\affiliation{Shanghai Branch, National Laboratory for Physical Sciences at Microscale and Department of Modern Physics, University of Science and Technology of China, Hefei, Anhui 230026, China}
\affiliation{CAS Center for Excellence and Synergetic Innovation Center in Quantum Information and Quantum Physics,
University of Science and Technology of China, Shanghai 201315, China}
\affiliation {CAS-Alibaba Quantum Computing Laboratory, Shanghai, 201315, China}

\begin{abstract}
Coherently manipulating a number of entangled qubits is the key task of quantum information processing.
In this article, we report on the experimental realization of a ten-photon Greenberger-Horne-Zeilinger state using thin BiB$_{3}$O$_{6}$ crystals.
The observed fidelity is $0.606\pm0.029$, demonstrating a genuine entanglement with a standard deviation of 3.6 $\sigma$.
This result is further verified using $p$-value calculation, obtaining an upper bound of $3.7\times10^{-3}$ under an assumed hypothesis test.
Our experiment paves a new way to efficiently engineer BiB$_{3}$O$_{6}$ crystal-based
multi-photon entanglement systems, which provides a promising platform
for investigating advanced optical quantum information processing tasks
such as boson sampling, quantum error correction and quantum-enhanced measurement.
\end{abstract}

\maketitle

\section{Introduction}
Quantum entanglement is fundamental to the field of quantum information processing
and to the broader foundations of quantum physics \cite{Nielsen2010}.
Over the course of the last few decades, numerous efforts have been devoted to
entanglement realization using various physical systems,
which include photons \cite{Pan12rmp}, ion traps \cite{Blatt08},
and superconducting qubits \cite{Clarke08}.
Being ideal carriers of quantum information,
photons are the main building blocks in the fields of
quantum communications  \cite{Gisin02,yuan10,Bouwmeester97,Yin12,Ma12,Luhe14,Lu16QSS},
quantum metrology \cite{Giovannetti04}, and quantum computing  \cite{KLM01,Kok07,Walther05,Kai07,Lu07shor,Lanyon07,Yao12TEC,Cai13,Spring13,Broome13,Tillmann13,Crespi13,Spagnolo14,Carolan15,Wang15}.
The experimental abilities to address and control a large number of entangled photons \cite{Bouwmeester99,Pan01prl,Zhao04,Lu07,Yao12,Huang11}
underpin the power of optical quantum technologies. For instance,
Aaronson and Arkhipov have predicted that given $\gtrsim$20 indistinguishable single photons,
boson sampling can reach a computational complexity intractable for classical computers \cite{Aaronson11}.

However, increasing the number of entangled photons in a given setup presents many challenges,
where despite significant improvements in developing experimental techniques that
generate multi-photon entangled states, the current record number of entangled photons is still eight,
in a system that has been demonstrated only recently \cite{Yao12,Huang11}.

When using spontaneous parametric down conversion (SPDC) \cite{Kwiat95} to create a large number of entangled photons,
it is crucial to increase the brightness of the entangled photon pairs.
This can be established by enhancing the photons’ collection efficiency $\xi$ rather than
increasing the total pair generation rate $R_{T}$ in order to suppress the contamination associated with double pair emission \cite{note1,Laskowski09}.
Note that the spatial walk-off resulting from the birefringence of SPDC crystals significantly influences $\xi$.
For a given SPDC crystal, a higher $\xi$ can be obtained by decreasing the walk-off of the SPDC photons in the crystal \cite{note7},
which can be realized by reducing the crystal’s length.
However, a thinner crystal can lead to a lower $R_{T}$ \cite{Ling08},
which implies that the observation of larger number of entangled photons is challenging,
even when thin type-II BBO crystals are employed.

In this study, BiB$_{3}$O$_{6}$ (BiBO) crystals are used to eliminate the spatial walk-off while maintaining a moderate $R_{T}$.
Comparing to the BBO crystals, BiBO crystals have a smaller spatial walk-off angle $\delta_{\theta}$
and higher type-II second-order nonlinear coefficient $d_{\text{eff}}^{\text{II}}$ \cite{note2}.
The generation of entangled photons through BiBO crystals has been reported,
which was achieved using the type-I \cite{Rangarajan09} or the type-II \cite{halevy11} SPDC process.
However, these techniques are not advanced enough for the realization of ten-photon entanglement.
Here, we present a technique for producing ultra-bright entangled photon pairs,
which relies on utilizing the Bell state synthesizer architecture \cite{Kim03} for thin BiBO crystals.
Our numerical calculations and experimental results demonstrate an enhanced $\xi$ when using thin BiBO crystals.
Consequently, this technique can be used to efficiently generate multi-photon entangled systems using type-II BiBO crystals.

The following sections provide a detailed description of our BiBO crystals based ten-photon entanglement system.
In Sec.~\ref{Sec:scheme}, a BiBO-based Bell state synthesizer is introduced.
Sec.~\ref{Sec:Implement} presents the experimental implementation of our ten-photon entanglement system.
The experimental results are presented in Sec.~\ref{Sec:results}.
In Sec.~\ref{Sec:conclusion} we provide a summary of this work and discuss its potential applications.
Further details regarding BiBO crystals and the $p$-value that concerns the readability of the research article are given in Sec.~\ref{Sec:method}.

\section{A SPDC source based on BiBO crystals}\label{Sec:scheme}

When compared with a BBO crystal, BiBO is expected to have a smaller $\delta_{\theta}$ and
a higher $d_{\text{eff}}^{\text{II}}$ (see Method. \ref{Sec:appBIBO}).
These two advantages indicate that a relatively thin BiBO structure can
yield a large $\xi$ with negligible effects on $R_{T}$, comparing to BBO crystals.
Since $\xi$ is inversely proportional to $L$, whereas $R_{T}$ is directly proportional to $L$ (see Method. \ref{Sec:appBIBO}),
selecting a suitable BiBO length (\textit{L}) is crucial to optimize the trade-off between $\xi$ and $R_{T}$.

\begin{figure}[htb]
\centering
\fbox{\includegraphics[width=0.925\linewidth]{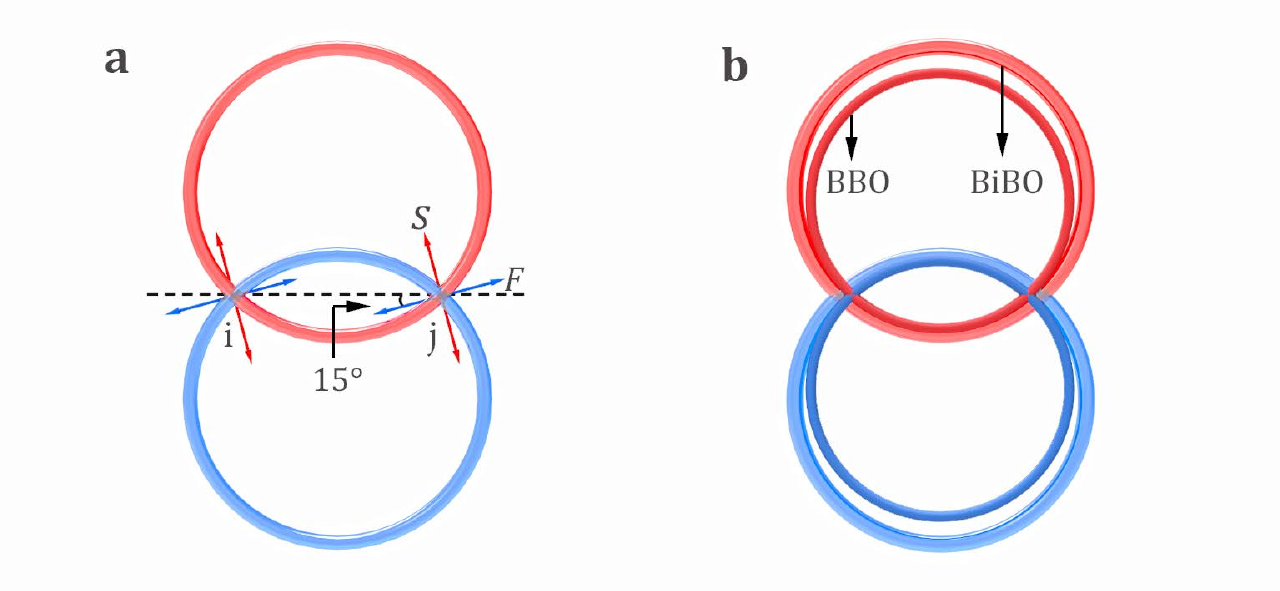}}
\caption{\textbf{Numerical simulations for the SPDC photon rings through a 3 nm bandpass filter.}
\textbf{a.} The different polarizations of the birefringent rays in a BiBO crystal.
The blue (red) ring represents the spatial distribution of the signal (idler) photons.
If the vector that connects the two intersections of the SPDC rings is parallel to $H$,
the $F$ ($S$) has a 15$^{\circ}$ deflection from $H$ ($V$) \cite{note3},
which can be calculated using the electric field vector \textbf{E}
for the 390~nm $\rightarrow$ 780~nm type-II SPDC process.
\textbf{b.} The respective SPDC photon rings of BBO and BiBO crystals.
The wave vector \textbf{k} is solely used to describe the spatial distributions with in the system.
In this simulation, the FWHM of the pump laser is assumed to be approximately 2.1 nm.
}
\label{fig:ten-sch}
\end{figure}

After comparing the parameters of the entangled photon pairs generated from the different SPDC crystals (see Method. \ref{Sec:appBIBO}),
we finally choose a 0.6 mm BiBO crystal to implement the Bell state synthesizer \cite{Kim03},
as illustrated in Fig. \ref{fig:ten-exp}b. In this setup, two birefringent compensators,
constituted by the first two half-wave plates (HWPs) and 0.3 mm BiBO crystals,
are used to eliminate the walk-off between the SPDC photons \cite{note8}.
As illustrated in Fig. \ref{fig:ten-sch}a, the polarizations of the SPDC photons emitted from the BiBO crystal are labeled as fast ($F$) and slow ($S$), respectively.
The last two HWPs are introduced to not only ensure the identical polarization for the SPDC photon pairs
when reaching the polarizing beam splitter (PBS) but also transform the polarizations $F$ and $S$ into vertical ($V$) and horizontal ($H$), respectively.
After interfering at the PBS, the SPDC photons with an (original) $F$ (blue) and an (original) $S$ polarization (red) are separated,
then detected by different single photon counting modules (SPCMs).
Note that this setup effectively disentangles the timing information from the polarization information for a given SPDC photon pair;
therefore, eliminating the need for spectral filtering.

As a biaxial crystal with a low degree of symmetry,
the BiBO crystals present many complicated properties.
For instance, as illustrated in Fig. \ref{fig:ten-sch}a, photons with identical polarizations at the two intersections of the SPDC rings, labeled as $i$ (left) and $j$ (right),
possess different full width at half maximum (FWHM) values (see Method. \ref{Sec:appBIBO}).
Besides, the final entangled photon pair generated from a non-collinear type-II BiBO crystal is not an ideal Bell state.
which can be expressed as $|\phi^{+}\rangle_{ij} = \textrm{cos} (7\pi/30)|HH\rangle _{ij} + \textrm{sin} (7\pi/30)|VV\rangle _{ij}$
according to our theoretical calculations (see Method. \ref{Sec:appBIBO}).

Fig. \ref{fig:ten-sch}b shows the respective spatial distributions of the SPDC photons generated from the BBO and BiBO crystals.
It's seen that the SPDC photons generated from the BiBO possess a larger divergence.
This is attributed to the difference in the dispersion value $\textrm{d}n/\textrm{d}\lambda$,
where this value is much larger in a BiBO crystal, as indicated in Ref. \cite{halevy11}.
Other detailed comparisons, involving $d_{\text{eff}}^{\text{II}}$ and $\delta_{\theta}$, can be found in Method. \ref{Sec:appBIBO}.

\section{Experiment implementation}\label{Sec:Implement}

\begin{figure*}[htb]
\centering
\fbox{\includegraphics[height=0.5\linewidth]{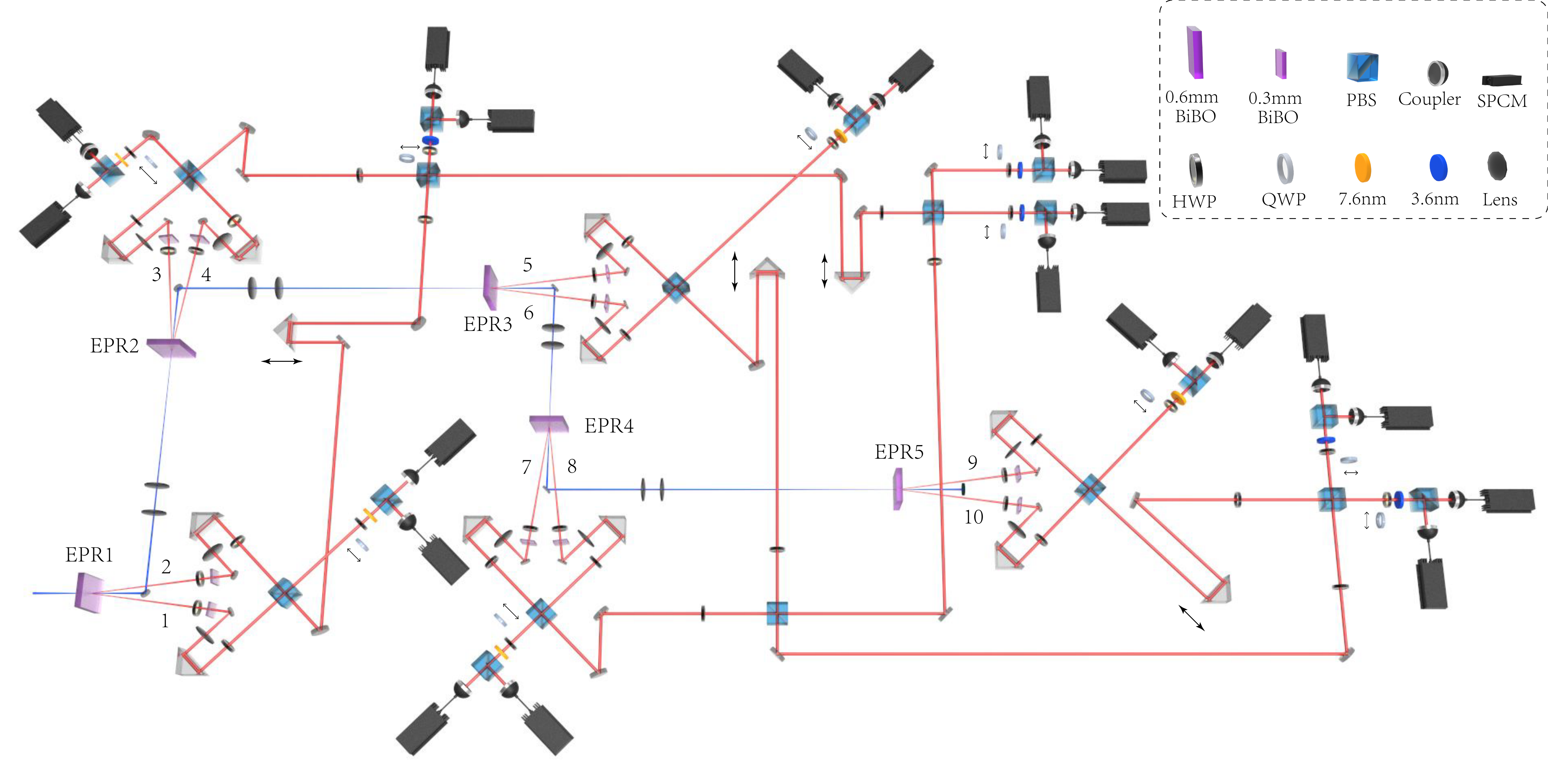}}
\caption{\textbf{Experimental setup for preparing ten photon GHZ state.}
An ultrafast pump laser with a central wavelength of 390~nm and a FWHM of 2.1~nm is successively sent through
the BiBO crystals to generate polarization-entangled photon pairs, i.e., EPR~1 $\sim$ EPR~5.
The distance between the 1st and 5th BiBO crystal is 2.65~meters.
In each BiBO-based Bell state synthesizer architecture, lenses with the focal length of 400~mm are placed to maximize the $\xi$.
The polarization of each output photon is analyzed using a combination of
a quarter-wave plate (QWP), a HWP and a PBS, together with a single-mode, fibre-coupled SPCM in each
output of the PBS.
Bandpass filters with $\Delta\lambda^{\mathrm{filter}\cdot\text{s}}_{\mathrm{FWHM}} = 3.6$~nm
on path 2, 3, 5, 7, 9 are used to erase the time information between the five entangled photon pairs \cite{Grice01}.
The other bandpass filters with $\Delta\lambda^{\mathrm{filter}\cdot\text{i}}_{\mathrm{FWHM}} = 7.8$~nm are chosen to achieve a maximum $\xi$.
We engineer these five entangled photon pairs into a ten-photon GHZ state by combining five signal photons
on a linear optical network consisting of four PBSs.
}
\label{fig:ten-exp}
\end{figure*}

In this experiment, we aimed to produce a ten-photon Greenberger-Horne-Zeilinger (GHZ) state, which can be expressed as
\begin{equation}
\ket{\mathrm{GHZ}_{10}} = \frac{1}{\sqrt{2}}\left(\ket{H}^{\otimes10}+\ket{V}^{\otimes10}\right).
\label{eq:GHZstate}
\end{equation}
The relevant experimental setup is shown in Fig.~\ref{fig:ten-exp}. Five independent entangled photon pairs were produced by sending an ultrafast laser
with a central wavelength of 390 nm through five 0.6 mm BiBO crystals.
A 1.05 W pump laser is focused onto each BiBO crystal with a beam waist of $\omega_{0}\simeq 85~\text{um}$,
to ensure having a suitable $R_{T}$. When the spectral filters are absent,
the typical twofold coincidence counting rate for each entangled photon pair is approximately 1,880,000~$\text{s}^{-1}$,
with an average $\xi=46.5\pm1\%$. In this case, the visibility \cite{Pan12rmp} in the $|D/A\rangle = (|H\rangle\pm|V\rangle)/\sqrt{2}$
and $|H/V\rangle$ basis is measured at 87.7$\%$ and 89.3$\%$, respectively.

We assume the SPDC photon with an original $F$ ($S$) polarization as signal (idler).
In our set-up, the average FWHM of the signal and idler photons is measured to be 7 nm and 14 nm, respectively.
Considering both $\xi$ and the coherence time of SPDC photons, bandpass filters with 3.6 nm and 7.8 nm are selected to
spectrally filter the signal and idler photons.
Eventually, the respective twofold coincidence counting rate of the five entangled photon pairs drops down to
605,000$\text{s}^{-1}$, 655,000$\text{s}^{-1}$, 590,000$\text{s}^{-1}$, 560,000$\text{s}^{-1}$, 515,000$\text{s}^{-1}$ \cite{note4},
with the corresponding $\xi$ measured at 37.3\%, 39.0\%, 37.0\%, 38.0\%, 36.8\%, respectively.
Thus, $\xi$ is relatively improved by $\sim$ 40\% when compared with the 2~mm BBO crystals \cite{Yao12}.
This makes the tenfold coincidence counting rate ($\sim R_{T} ^{5}\xi ^{10}$/16) increase to approximately 0.5 counts per hour,
which is 27 times higher than the case when we directly adopt the techniques
in Ref. \cite{Yao12} to demonstrate the ten-photon entanglement.


Next, the signal photons (paths 2, 3, 5, 7, and 9) are directed to the PBSs to
ensure the spatial indistinguishability between photons from the different SPDC sources.
Through fine adjustment, the photons simultaneously arrive at the PBSs, resulting in an average visibility of 71.5\%,
a value that is obtained when photons experience a Hong-Ou-Mandel-type interference \cite{HOM87} at four PBSs.

Since each entangled photon pairs from the BiBO crystals is an imperfect Bell state,
the polarization of SPDC photons from the 4th and 5th BiBO crystals is rotated by $90^{\circ}$,
which would transform the prepared two photon entangled state to
$|\phi^{+}\rangle_{ij}^{'} = \textrm{cos} (7\pi/30)|VV\rangle _{ij} + \textrm{sin} (7\pi/30)|HH\rangle _{ij}$
to minimize the unbalance between the final $|H\rangle ^{\otimes10}$ and $|V\rangle ^{\otimes10}$ components.
In this case, our final ten-photon entangled state can be formulated theoretically as
$|\Phi^{+}\rangle = \textrm{cos} (7\pi/30)|H\rangle ^{\otimes10} + \textrm{sin} (7\pi/30)|V\rangle ^{\otimes10}$.

\section{Result}\label{Sec:results}

\begin{figure*}[htb]
\centering
\fbox{\includegraphics[width=0.95\linewidth]{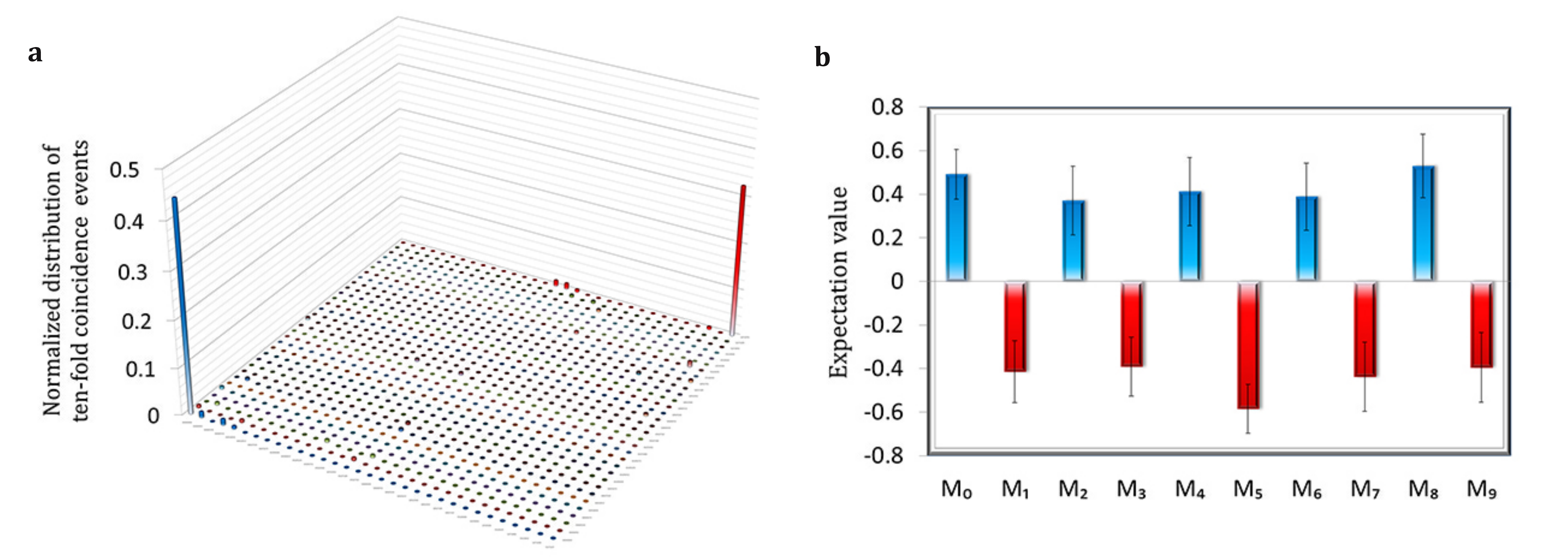}}
\caption{\textbf{Experimental results for the ten-photon GHZ state.}
\textbf{a.} The population of the prepared tenfold coincidence events in the $H/V$ basis.
The total measured time is 300 h.
\textbf{b.} Expectation values in the basis of $M_k^{\otimes 10}$, $k=0,1,\dots,9$.
The $M_0 (\sigma_{x})$ and $M_5 (\sigma_{y})$ values are measured respectively for 110 h,
while the remaining eight observables are measured for 80 h.
Error bars indicate one standard deviation deduced from propagated Poissonian
counting statistics of the raw detection events.}
\label{fig:ten-data}
\end{figure*}

We measure the fidelity of our prepared ten-photon state to show the existence of genuine entanglement.
For an $n$-qubit GHZ state, one can have the following decomposition \cite{Otfried07}:
\begin{align}
\hat F& = \Ket{\text{GHZ}_n}\Bra{\text{GHZ}_n}\notag \\
& = \frac{1}{2}\left (\Ket{H}^{\otimes n}+\Ket{V}^{\otimes n}\right)\left (\Bra{H}^{\otimes n}+\Bra{V}^{\otimes n}\right) \notag \\
& = \sum_{k = 0}^{n-1}\alpha_kM_k^{\otimes n}+\frac{1}{2}\left ( (\Ket{H}\Bra{H})^{\otimes n}+ (\Ket{V}\Bra{V})^{\otimes n}\right),
\label{eq:proj_ghz}
\end{align}
where $\alpha_k = (-1)^k/ (2n)$ and $M_k = \cos (k\pi/n)\sigma_x + \sin (k\pi/n)\sigma_y$,
$k = 0,1,...,n-1$. Hence, to estimate the fidelity $\bar F = \mathrm{tr}[\hat F\rho_{n}]$ of the prepared state $\rho_{n}$,
one can measure the correlations
under local measurement settings $M_k^{\otimes n}$, $k = 0,1,...,n-1$, and also the
probabilities of $ (\sigma_{z,1},\sigma_{z,2},...,\sigma_{z,n}) = (H,H,...,H)$ and $ (V,V,...,V)$ in the $H/V$ basis.
In experiment, the fidelity can be estimated by
\begin{equation}
\bar F = \sum_{k = 0}^{n-1}\alpha_k \frac{N_k^{+}-N_k^{-}}{N_k} + \frac{1}{2}\frac{N_z^{0}+N_z^{1}}{N_z}.
\label{eq:exp_est}
\end{equation}
where $N_k^{+}$ ($N_k^{-}$) is the number of trials with positive (negative) correlation
under measurement setting $M_k^{\otimes n}$, $k = 0,1,...,n-1$, and $N_z^{0}$  ($N_z^{1}$)
is the number of trials with outcomes $ (\sigma_{z,1},\sigma_{z,2},...,\sigma_{z,n}) = (H,H,...,H)$ ($ (V,V,...,V)$).

In our experiment, we post-select the tenfold coincidence counting events,
in which only one SPCM on each path registers, as a valid experimental data.
Eventually, a complete set of the 1024 tenfold coincidence events
are simultaneously registered for entanglement verification by a homemade FPGA-based coincidence unit.
All the 1024 polarization distributions in the $H/V$ basis are illustrated in Fig. \ref{fig:ten-data}a,
from which we can see that $|H\rangle^{\otimes10}$ and $|V\rangle^{\otimes10}$ are the dominant parts
in the overall tenfold coincidence events. This demonstrates a total signal-to-noise ratio of 3.36:1.
Furthermore, measurements in the $M_{k}^{\otimes 10} = [\cos (k\pi/10)\sigma_x + \sin (k\pi/10)\sigma_y]^{\otimes10}$, $k = 0,1,...,9$ basis
are performed to verify whether the $|H\rangle^{\otimes10}$ and $|V\rangle^{\otimes10}$ components are in coherent superposition,
yielding an average signal-to-noise ratio of 2.58:1. The expectation values for each $M_{k}^{\otimes 10}$ are illustrated in Fig. \ref{fig:ten-data}b.
Note that the average visibility in $M_{k}^{\otimes 10}$ (= 0.442 $\pm$ 0.046) is lower than that in the case of $H/V$ (= $0.542\pm0.070$) polarization.
This is attributed to the unbalance between the $|H\rangle ^{\otimes10}$ and $|V\rangle ^{\otimes10}$ and
the partial distinguishability of the signal photons from the different SPDC sources.
Given the aforementioned experimental results and Eq. (\ref{eq:exp_est}),
the calculated fidelity of our ten-photon GHZ state is
$\bar F_{\text{exp}} = 0.606\pm0.029$.
Ref.~\cite{Toth05} shows that the prepared multi-particle state is genuinely entangled
as long as the average $\bar F$ value is larger than 0.5.
Therefore, our experiment implements and proves the existence of a genuine ten-photon entanglement sate,
with a 3.6~$\sigma$ violation, based on Poisson’s statistics hypothesis.

Furthermore, we characterize the effect of statistical fluctuation within finite data without the Poisson-distribution assumption.
For any bi-separable state $\rho_{\rm{bs}}$ that satisfies ${F}_{\text{bs}} = \text{Tr} (\rho_\text{bs} \hat F)\leq 0.5$,
one can predict an estimated fidelity higher than or equal to the observed one $\bar F_{\mathrm{exp}}$ with non-zero probability.
This probability is called a $p$-value,
which determines the operational meaning of the experimental result in the hypothesis test of bi-seperable states \cite{Zhang11}.
With a small enough $p$-value, we can conclude that the experimental result is significantly incompatible with
any bi-separable state.  With the data observed in our experiment, the $p$-value is upper bound by $3.7\times10^{-3}$ (see Method. \ref{Sec:appvalue}).
In the analysis of estimating the standard deviation of $\bar F_{\mathrm{exp}}$,
we assume the experiment data to be independent and identically distributed.
It is worth mentioning that the estimation of $p$-value is free of such assumptions.

\section{Conclusion}\label{Sec:conclusion}
In summary, we have demonstrated the successful generation and characterization of a ten-photon GHZ state using a thin BiBO crystal.
By utilizing the entanglement witness, a genuine ten-photon entanglement with a 0.606 fidelity is demonstrated
with a standard deviation of 3.6 $\sigma$ and a $p$-value of $3.7\times10^{-3}$.
This work paves the way for multi-photon manipulation using thin non-linear crystals that simultaneously provide high $R_{T}$ and $\xi$ values,
allowing us to tackle new challenges in the field of optical quantum technology.
For instance, minor modifications can be conducted to our experimental setup, to achieve a quantum error correction code \cite{Laflamme96},
which is one of quantum computation’s long sought goals.
Another immediate application for our setup is boson sampling with numerous photons.
Further study of BiBO-based entangled photon pairs can be focused on the sandwich structure using beamlike type-II BiBO crystals.
As the signal-idler photon pairs are emitted into two separate circular beams instead of two diverging cones of (non-)collinear type-II SPDC,
a greater $\xi$ would be expected in the beamlike BiBO crystals.
Furthermore, the imperfection of the output entangled photon state from non-collinear type-II BiBO crystals can also be eliminated
by the aid of beamlike BiBO-based sandwich structure.
Recently, another ten-photon work with a fidelity of $0.573\pm0.023$ is reported in Ref. \cite{Wang16arxiv},
using beamlike BBO-based sandwich structure \cite{Takeuchi01}.
Combining the techniques present in these two ten-photon works,
one could expect a further improvement of $\xi$ using the beamlike BiBO-based sandwich structure.

\section{Method}\label{Sec:method}
\subsection{Details for BiBO crystal}\label{Sec:appBIBO}
This section provides a theoretical description of type-II BiBO (BBO) phase-matching of 390 nm $\rightarrow$ 780 nm SPDC.
For BiBO and BBO crystals, the maximal collinear $d_{\text{eff}}^{\text{II}}$ is calculated to be $1.94~\text{pm/V}$ and $1.15~\text{pm/V}$ \cite{note5}, respectively.
Considering the 390 nm $\rightarrow$ 780 nm non-collinear type-II phase-matching condition,
the populations of emitted SPDC photon pairs are unbalanced owing to the low symmetry of BiBO crystals.
For clarity, the photon that propagates from the left (right) intersection in Fig. \ref{fig:ten-sch}a is labeled as $i$ ($j$).
When choosing the non-collinear type-II phase-matching angle present in Ref. \cite{halevy11},
the $d_{\text{eff}}^{\text{II}}$ of the $|FS\rangle_{ij}$ and $|SF\rangle_{ij}$ components are calculated to be 1.84~pm/V and 2.02~pm/V, respectively.
Consequently, the resulting two-photon entangled state can be written as:
$|\phi^{+}\rangle_{ij} = \textrm{cos} (7\pi/30)|HH\rangle _{ij} + \textrm{sin} (7\pi/30)|VV\rangle _{ij}$.

We further calculate the walk-off angle $\delta_{\theta}$ for the BiBO and BBO crystals.
For simplicity, $\delta_{\theta}$ is considered to account solely for the SPDC rays.
BBO is a uniaxial crystal where the walk-off only occurs for the SPDC photons that have an extraordinary (e) polarization.
This walk-off is calculated to be $\delta_{\theta}^{\text{BBO}} = 0.072~\text{rad}$.
However, the BiBO crystal has a more complex biaxial symmetry,
where both of the down-converted photons have respective spatial walk-off values of
0.020~\text{rad} and 0.063~\text{rad}.
Nevertheless, the overall spatial walk-off magnitude in a BiBO crystal is estimated to be equal to
$\delta_{\theta}^{\text{BiBO}} = 0.066~\text{rad}$, which is smaller than $\delta_{\theta}^{\text{BBO}} $.

To determine a suitable crystal length $L$ value for the BiBO crystal, four crystals are tested: 2~mm BBO, 1~mm BBO, 1.2~mm BiBO and 0.6~mm BiBO.
In the test, a 920 mW pump power with a pump beam waist of $\sim90$ $\mu$m is used.
This test reveals the different relationships among collection efficiency $\xi$, total pair generation rate $R_{T}$ and $L$.
For comparison, all the experiment results are relative values with respect to those of a 2~mm BBO.

Tables \ref{tab:efficiency} shows the relationship between $\xi$ and walk-off values.
The FWHM values of SPDC photons are also given for different crystals and
$\xi$ are measured without using the bandpass filters.
According to Tables \ref{tab:efficiency}, it can be seen that the increase of $\xi$ is inversely proportional to the decrease of walk-off value.
Especially, the $\xi$ of 0.6~mm BiBO has relatively increased by $\sim$ 42.6\% comparing to that of 2~mm BBO.

\begin{table}[htb]
\centering
\caption{\bf The experimental relationship between the increase of $\xi$ and decrease of the spatial walk-off value.}
\begin{tabular}{cccc}
\hline
crystal  & walk-off $\Downarrow$   & $\xi \Uparrow$ & FWHM\\
\hline
2~mm BBO  & 0  & 0 & 7.5~nm ($e$), 15.5~nm ($o$) \\
\hline
1~mm BBO  & 54.1$\%$  & 30.2$\%$ & 9.6~nm ($e$), 15.5~nm ($o$)\\
\hline
\multirow{2}*{1.2~mm BiBO}  & \multirow{2}*{53.8$\%$}  & \multirow{2}*{30.0$\%$} & \small 5.8~nm ($F_{1}$), 15.6~nm ($S_{1}$)\\
                            &                          &                         & \small 5.5~nm ($F_{2}$), 15.6~nm ($S_{2}$)\\
\hline
\multirow{2}*{0.6~mm BiBO}  & \multirow{2}*{72.5$\%$}  & \multirow{2}*{42.6$\%$} & \small 6.8~nm ($F_{1}$), 17.5~nm ($S_{1}$)\\
                            &                          &                         & \small 7.3~nm ($F_{2}$), 16.2~nm ($S_{2}$)\\
\hline
\end{tabular}
  \label{tab:efficiency}
\end{table}

\begin{table}[htb]
\centering
\caption{{\bf The experimental relative $R_{T}$ under different bandpass filters.}
$R_{T}^{3, 3}$, $R_{T}^{3, 8}$, $R_{T}^{\infty, \infty}$, $R_{T}^{3.6, 7.8}$ is the corresponding
experimental relative $R_{T}$ under different bandpass filters configuration of (3 nm, 3 nm), (3 nm, 8nm), (no filters, no filters), (3.6 nm, 7.8nm), respectively.
The $R_{T}^{3.6, 7.8} = 0.488$ of 0.6 mm BiBO crystals is calculated with respect to $R_{T}^{3, 8}$ of 2 mm BBO.}
\begin{tabular}{ccccc}
\hline
crystal  & $R_{T}^{3, 3}$    & $R_{T}^{3, 8}$ & $R_{T}^{\infty, \infty}$  &$R_{T}^{3.6, 7.8}$\\
\hline
2~mm BBO  & 1  & 1 & 1 & -\\
\hline
1~mm BBO  & 0.517  & 0.569 & 0.455 & -\\
\hline
1.2~mm BiBO  & 0.796  & 0.859 & 0.764 & -\\
\hline
0.6~mm BiBO  & 0.449  & 0.483 & 0.413 & 0.488\\
\hline
\end{tabular}
  \label{tab:RT}
\end{table}

The experimental relative $R_{T}$ of the four crystals under different bandpass filters are shown in Table. \ref{tab:RT}.
For a given crystal (BiBO or BBO), one can conclude that $R_{T}$ is nearly proportional to $L$ in all of our bandpass filters configuration.
However, when referring to different crystals, e.g., between BiBO and BBO,
the relationship is not that clear since $R_{T}$ is not simply determined by $d_{\text{eff}}$ and $L$.
According to the Table \ref{tab:RT}, for the same $L$, the $R_{T}$ of BiBO crystals can be relatively enhanced by 40\% $\sim$ 50\%
with respect to the $R_{T}$ of BBO.
To verify the experimental results,
we perform theoretical calculations of $R_{T}^{\infty, \infty}$ \cite{Ling08} using
\begin{equation}\label{}
  R_{T}^{\infty, \infty} = (\frac{d_{\text{eff}}^{\text{BiBO}}}{d_{\text{eff}}^{\text{BBO}}})^2\cdot\frac{L_{\text{BiBO}}}{L_{\text{BBO}}}\cdot\frac{[n_{p}n_{s}n_{i}(n_{i}-n_{s})]_{\text{BBO}}}{[n_{p}n_{s}n_{i}(n_{i}-n_{s})]_{\text{BiBO}}}\cdot\frac{\Omega_{\text{BiBO}}}{\Omega_{\text{BBO}}}.
\end{equation}
Here, $n_{p}$, $n_{s}$ and $n_{i}$ represent the refractive indices of the pump, signal, idler lights, respectively.
The spectral integral $\Omega$ depends on the walk-off parameter $\Delta$ \cite{Ling08}.
We first calculate the $n_{p}$, $n_{s}$, $n_{i}$ and $\Delta$ values, which are shown in Table \ref{tab:RT_cal}.
Then by substituting these values, we obtain a theoretical $R_{T}^{\infty, \infty}$ of 0.424 for 0.6 mm BiBO crystals,
which agrees with our experimental result of 0.413.

\begin{table}[htb]
\centering
\caption{\bf Theoretical values of $n_{p}$, $n_{s}$, $n_{i}$ and $\Delta$ in BBO and BiBO crystals}
\begin{tabular}{ccccc}
\hline
crystal  & $n_{p}$    & $n_{s}$ & $n_{i}$  &$\Delta$\\
\hline
2~mm BBO  & 1.63  & 1.60 & 1.66 & 0.82\\
\hline
0.6~mm BiBO  & 1.84  & 1.78 & 1.90 & 0.28\\
\hline
\end{tabular}
  \label{tab:RT_cal}
\end{table}

Moreover, we perform some theoretical simulations, such as the collinear type-II phase-matching angles,
$d_{\text{eff}}^{\text{II}}$ and the spatial walk-offs \cite{note6}.
It's remarkable that around the collinear type-II phase-matching region with minimal spatial walk-off angle ($\sim$ 0.011~rad) of BiBO crystals,
$d_{\text{eff}}^{\text{II}}$ is calculated to be 1.1~\text{pm/V},
almost the same with that of BBO crystals.
This region may offer opportunities to create entangled photon pairs with even higher $\xi$ since it has extremely small walk-off value.

\begin{figure}[htb]
\centering
\includegraphics[width=\linewidth]{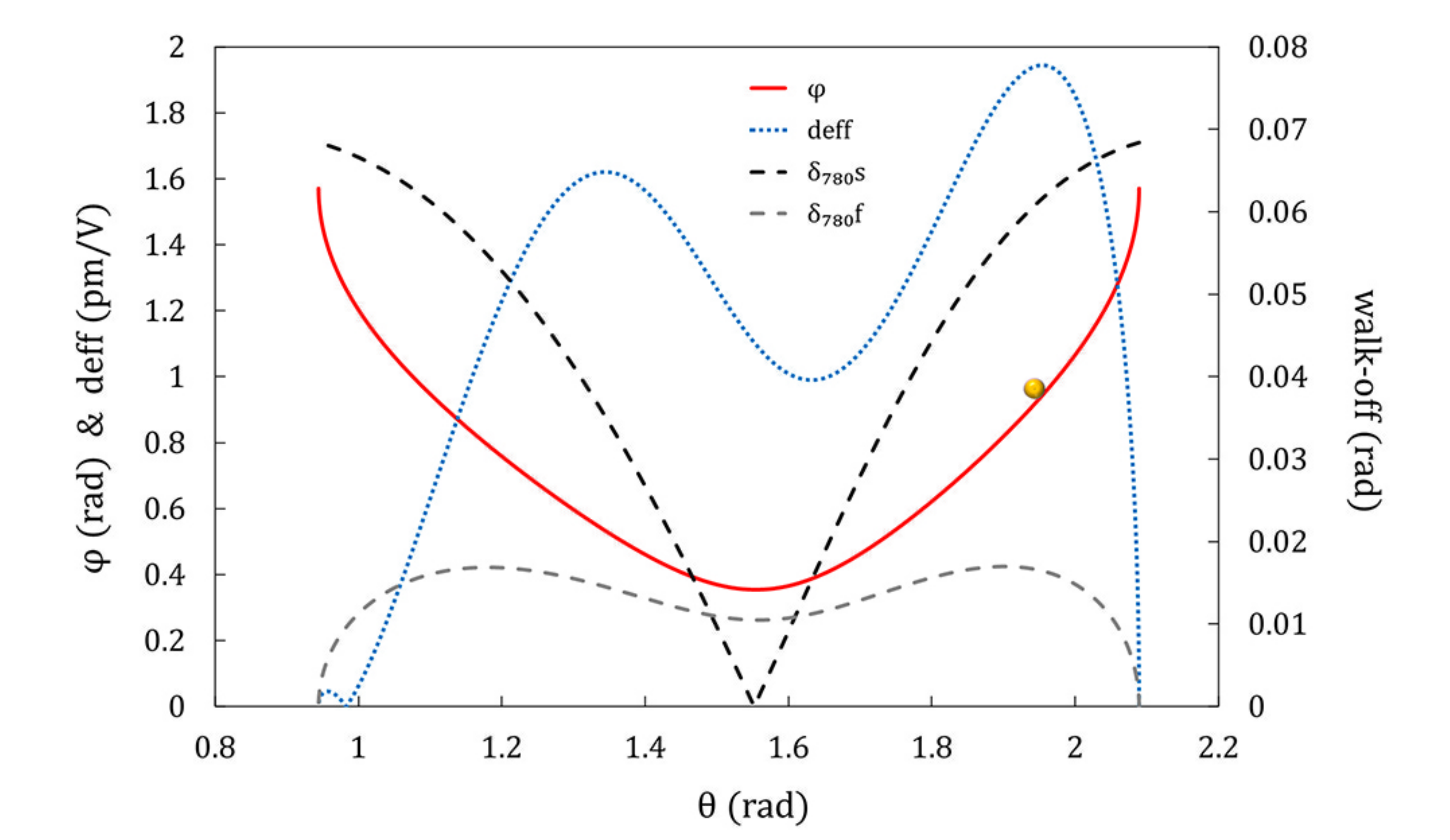}
\caption{\textbf{Theoretical simulation curves of the collinear type-II phase-matching angles,
$d_{\text{eff}}^{\text{II}}$ and the spatial walk-off for a BiBO crystal.}
The collinear type-II phase-matching angles ($\theta$, $\varphi$)
in the main refractive indices coordinates (solid red), $d_{\text{eff}}^{\text{II}}$ (dotted blue),
$\delta_{\theta}$ of 780~nm slow (dashed black) and fast (dashed grey) photons are simulated, respectively.
The yellow dot represents the non-collinear type-II phase-matching angle (1.944~rad, 0.962~rad) used in the experiment.
}
\label{fig:bibo}
\end{figure}

\subsection{Estimation of the $p$-value}\label{Sec:appvalue}
Due to statistical fluctuations in finite number of data points, it is possible
that a bi-separable state can predict a fidelity no less than the observed fidelity with non-zero probability.
This probability is called
a $p$-value, which determines the operational meaning of the experimental result~\cite{Zhang11}.
To bound the $p$-value, we can think the experiment as a hypothesis test 
of the inequality $\text{Tr}(\rho_\text{bs} \hat F)\leq F_0 = 0.5$, which is satisfied
by all bi-separable states $\rho_{\rm{bs}}$.
We assume that the measurement outcomes from different trials are independent but not necessarily identical.
In each trial of the test, one selects a measurement setting and records an outcome.
When measurement $\sigma_k^{\otimes n}$ is performed,
the trial outcome is $\pm\alpha_kN_t/{N_k}$,
where $\alpha_k= (-1)^k/ (2n)$ and the sign $\pm$ depends on whether the observed correlation is positive or
negative. Here, $N_t=(N_z + \sum_{k=0}^{n-1}N_k)$ is the total number of trials in the test and
$N_z$ and $N_k$ denote the coincidence counts in the $H/V$  and $M_k$ bases, respectively.
When measurement $\sigma_z^{\otimes n}$ is performed,
the trial outcome is ${N_t}/{2N_z}$ or $0$, depending on whether the measurement outcome
is $(\sigma_{z,1},\sigma_{z,2},...,\sigma_{z,n})=(0,0,...,0)$/$(1,1,...,1)$ or not.

Denote the value for the $i$th trial, for $i = 1, 2, \dots, N_t$, by $F_i$, then the averaged experiment estimation is
\begin{equation}\label{}
  \bar F_{\mathrm{est}} = \frac{1}{N_t}\sum_{i=1}^{N_t} F_i.
\end{equation}
It is straightforward to see that this estimation is the same as the one of fidelity $\bar F$ according to Equation \ref{eq:exp_est}.

For bi-separable state $\rho_{\mathrm{bs}}$, the average of $F_i$ is smaller than $F_0 = 0.5$ \cite{Toth05}.
Hence, by denoting a sequence $F^K_{\mathrm{bs}}$ by
\begin{equation}\label{}
   F^K_{\mathrm{bs}} = \sum_{i=1}^{K} (F_i-F_0),
\end{equation}
we can easily prove that the sequence of $F^K_{\mathrm{bs}}$ is a super-martingale and $\bar F_{\mathrm{bs}} = F^{N_t}_{\mathrm{bs}}/N_t$.
For such a super-martingale sequence, the $p$-value that the estimation $\bar F_{\mathrm{bs}}$ achieves an observed value
$\bar F_{\mathrm{exp}}$ can be bounded according to  Corollary 2.2 of Pinelis's paper~\cite{Pinelis06}
\begin{equation}\label{eq:p-bound4}
  p = \text{Prob}_\text{bs}(\bar F_{\mathrm{bs}} \geq\bar F_{\mathrm{exp}})\leq D\left( \frac{N_t (\bar F_{\mathrm{exp}}-F_0)}{S_{N_t}} \right),
\end{equation}
where the function $D(x) = \min\{\text{exp}(-x^2/2), 5!(e/5)^5 I(x)\}$ and the function $I(x)$
is the cumulative tail distribution function of the standard normal distribution.
Here, $S_{N_t} = (s_1^2 + s_2^2 + \dots+s^2_{N_t})^{1/2}$ and $s_i = (\max F_i - \min F_i)/2$, for $i = 1, 2, \dots, N_t$.
For the $i$th trial, we have $s_i = N_t/(4N_z)$ and $s_i = \alpha_kN_t/(N_k)$ when the $H/V$ basis and the $M_k$ basis are chosen, respectively.

According to the definition of fidelity $\bar F$ in Equation \ref{eq:exp_est}, $S_{N_t}$ is given by
\begin{equation}\label{}
\begin{aligned}
  S_{N_t} & = \sqrt{\sum_{i=1}^{N_t}s_i^2},\\
& = \sqrt{\left(\frac{N_t}{4N_z}\right)^2\times N_z + \sum_{k=0}^{9} \left(\frac{\alpha_kN_t}{N_k}\right)^2\times N_k},\\
& = N_t\sqrt{\frac{1}{16N_z} + \sum_{k=0}^{9}\frac{\alpha_k^2}{N_k}}.
\end{aligned}
\end{equation}
Therefore, the $p$-value can be upper bounded by
\begin{equation}\label{}
  p = \text{Prob}_\text{bs}(\bar F_{\mathrm{bs}} \geq\bar F_{\mathrm{exp}})\le D\left( \frac{
  (\bar F_{\mathrm{exp}}-F_0)}{\sqrt{\frac{1}{16N_z}+\sum_{k=0}^{9}\frac{\alpha_k^2}{N_k}}} \right),
\end{equation}
In experiment, we have an observed average fidelity $\bar F_{\mathrm{exp}} = 0.606$ and the values of $N_z$ and $N_k$ summarized in Table \ref{tab:s}.
With our experiment results, we calculate the upper bound of the $p$-value to be $p\le 3.7\times 10^{-3}$.

The inequality of Equation \ref{eq:p-bound4} reads as that, the probability according to any bi-separable
state of predicting a fidelity $\bar{F}_{\text{bs}}$ in the experiment not
lower than the observed fidelity $\bar{F}_{\mathrm{exp}}$ is not bigger than the $p$-value.
In other words, the confidence that a genuine multipartite
entangled state is prepared, given the observed results, is at least as high as $1 - p$.

\begin{table}[htb]
  \centering
  \caption{\bf Tenfold coincidence counting events in the $H/V$ and $M_{k}^{\otimes 10}$ bases}
  \begin{tabular}{ccccccccccc}
    \hline
    $N_z$ & $N_{0}$& $N_1$ & $N_2$& $N_3$& $N_4$& $N_5$& $N_6$& $N_7$& $N_8$& $N_9$\\
    \hline
    144 & 53 & 34 & 46 & 35 & 41 & 59 & 33 & 34 & 32 & 36\\
    \hline
  \end{tabular}
  \label{tab:s}
\end{table}


\section*{Funding Information}
National Basic Research Program of China Grants No.~2011CB921300, No.~2013CB336800,
the National Natural Science Foundation of China Grants, the Chinese Academy of Sciences.

\section*{Acknowledgments}
The authors acknowledge insightful discussions with J. Zhang, J.-Y Fan, P. Xu, W.-H Jiang and N. Zhou.

\clearpage



\begin{thebibliography}{57}%
\makeatletter
\providecommand \@ifxundefined [1]{%
 \@ifx{#1\undefined}
}%
\providecommand \@ifnum [1]{%
 \ifnum #1\expandafter \@firstoftwo
 \else \expandafter \@secondoftwo
 \fi
}%
\providecommand \@ifx [1]{%
 \ifx #1\expandafter \@firstoftwo
 \else \expandafter \@secondoftwo
 \fi
}%
\providecommand \natexlab [1]{#1}%
\providecommand \enquote  [1]{``#1''}%
\providecommand \bibnamefont  [1]{#1}%
\providecommand \bibfnamefont [1]{#1}%
\providecommand \citenamefont [1]{#1}%
\providecommand \href@noop [0]{\@secondoftwo}%
\providecommand \href [0]{\begingroup \@sanitize@url \@href}%
\providecommand \@href[1]{\@@startlink{#1}\@@href}%
\providecommand \@@href[1]{\endgroup#1\@@endlink}%
\providecommand \@sanitize@url [0]{\catcode `\\12\catcode `\$12\catcode
  `\&12\catcode `\#12\catcode `\^12\catcode `\_12\catcode `\%12\relax}%
\providecommand \@@startlink[1]{}%
\providecommand \@@endlink[0]{}%
\providecommand \url  [0]{\begingroup\@sanitize@url \@url }%
\providecommand \@url [1]{\endgroup\@href {#1}{\urlprefix }}%
\providecommand \urlprefix  [0]{URL }%
\providecommand \Eprint [0]{\href }%
\providecommand \doibase [0]{http://dx.doi.org/}%
\providecommand \selectlanguage [0]{\@gobble}%
\providecommand \bibinfo  [0]{\@secondoftwo}%
\providecommand \bibfield  [0]{\@secondoftwo}%
\providecommand \translation [1]{[#1]}%
\providecommand \BibitemOpen [0]{}%
\providecommand \bibitemStop [0]{}%
\providecommand \bibitemNoStop [0]{.\EOS\space}%
\providecommand \EOS [0]{\spacefactor3000\relax}%
\providecommand \BibitemShut  [1]{\csname bibitem#1\endcsname}%
\let\auto@bib@innerbib\@empty
\bibitem [{\citenamefont {Nielsen}\ and\ \citenamefont
  {Chuang}(2000)}]{Nielsen2010}%
  \BibitemOpen
  \bibfield  {author} {\bibinfo {author} {\bibfnamefont {M.~A.}\ \bibnamefont
  {Nielsen}}\ and\ \bibinfo {author} {\bibfnamefont {I.~L.}\ \bibnamefont
  {Chuang}},\ }\href@noop {} {\emph {\bibinfo {title} {Quantum Computation and
  Quantum Information}}}\ (\bibinfo  {publisher} {Cambridge University Press},\
  \bibinfo {address} {Cambridge U.K.},\ \bibinfo {year} {2000})\BibitemShut
  {NoStop}%
\bibitem [{\citenamefont {Pan}\ \emph {et~al.}(2012)\citenamefont {Pan},
  \citenamefont {Chen}, \citenamefont {Lu}, \citenamefont {Weinfurter},
  \citenamefont {Zeilinger},\ and\ \citenamefont {\ifmmode~\dot{Z}\else
  \.{Z}\fi{}ukowski}}]{Pan12rmp}%
  \BibitemOpen
  \bibfield  {author} {\bibinfo {author} {\bibfnamefont {J.-W.}\ \bibnamefont
  {Pan}}, \bibinfo {author} {\bibfnamefont {Z.-B.}\ \bibnamefont {Chen}},
  \bibinfo {author} {\bibfnamefont {C.-Y.}\ \bibnamefont {Lu}}, \bibinfo
  {author} {\bibfnamefont {H.}~\bibnamefont {Weinfurter}}, \bibinfo {author}
  {\bibfnamefont {A.}~\bibnamefont {Zeilinger}}, \ and\ \bibinfo {author}
  {\bibfnamefont {M.}~\bibnamefont {\ifmmode~\dot{Z}\else \.{Z}\fi{}ukowski}},\
  }\href@noop {} {\bibfield  {journal} {\bibinfo  {journal} {Rev. Mod. Phys.}\
  }\textbf {\bibinfo {volume} {84}},\ \bibinfo {pages} {777} (\bibinfo {year}
  {2012})}\BibitemShut {NoStop}%
\bibitem [{\citenamefont {Blatt}\ and\ \citenamefont
  {Wineland}(2008)}]{Blatt08}%
  \BibitemOpen
  \bibfield  {author} {\bibinfo {author} {\bibfnamefont {R.}~\bibnamefont
  {Blatt}}\ and\ \bibinfo {author} {\bibfnamefont {D.}~\bibnamefont
  {Wineland}},\ }\href@noop {} {\bibfield  {journal} {\bibinfo  {journal}
  {Nature}\ }\textbf {\bibinfo {volume} {453}},\ \bibinfo {pages} {1008}
  (\bibinfo {year} {2008})}\BibitemShut {NoStop}%
\bibitem [{\citenamefont {Clarke}\ and\ \citenamefont
  {Wilhelm}(2008)}]{Clarke08}%
  \BibitemOpen
  \bibfield  {author} {\bibinfo {author} {\bibfnamefont {J.}~\bibnamefont
  {Clarke}}\ and\ \bibinfo {author} {\bibfnamefont {F.~K.}\ \bibnamefont
  {Wilhelm}},\ }\href@noop {} {\bibfield  {journal} {\bibinfo  {journal}
  {Nature}\ }\textbf {\bibinfo {volume} {453}},\ \bibinfo {pages} {1031}
  (\bibinfo {year} {2008})}\BibitemShut {NoStop}%
\bibitem [{\citenamefont {Gisin}\ \emph {et~al.}(2002)\citenamefont {Gisin},
  \citenamefont {Ribordy}, \citenamefont {Tittel},\ and\ \citenamefont
  {Zbinden}}]{Gisin02}%
  \BibitemOpen
  \bibfield  {author} {\bibinfo {author} {\bibfnamefont {N.}~\bibnamefont
  {Gisin}}, \bibinfo {author} {\bibfnamefont {G.}~\bibnamefont {Ribordy}},
  \bibinfo {author} {\bibfnamefont {W.}~\bibnamefont {Tittel}}, \ and\ \bibinfo
  {author} {\bibfnamefont {H.}~\bibnamefont {Zbinden}},\ }\href@noop {}
  {\bibfield  {journal} {\bibinfo  {journal} {Rev. Mod. Phys.}\ }\textbf
  {\bibinfo {volume} {74}},\ \bibinfo {pages} {145} (\bibinfo {year}
  {2002})}\BibitemShut {NoStop}%
\bibitem [{\citenamefont {Yuan}\ \emph {et~al.}(2010)\citenamefont {Yuan},
  \citenamefont {Bao}, \citenamefont {Lu}, \citenamefont {Zhang}, \citenamefont
  {Peng},\ and\ \citenamefont {Pan}}]{yuan10}%
  \BibitemOpen
  \bibfield  {author} {\bibinfo {author} {\bibfnamefont {Z.-S.}\ \bibnamefont
  {Yuan}}, \bibinfo {author} {\bibfnamefont {X.-H.}\ \bibnamefont {Bao}},
  \bibinfo {author} {\bibfnamefont {C.-Y.}\ \bibnamefont {Lu}}, \bibinfo
  {author} {\bibfnamefont {J.}~\bibnamefont {Zhang}}, \bibinfo {author}
  {\bibfnamefont {C.-Z.}\ \bibnamefont {Peng}}, \ and\ \bibinfo {author}
  {\bibfnamefont {J.-W.}\ \bibnamefont {Pan}},\ }\href@noop {} {\bibfield
  {journal} {\bibinfo  {journal} {Phys. Rep.}\ }\textbf {\bibinfo {volume}
  {497}},\ \bibinfo {pages} {1} (\bibinfo {year} {2010})}\BibitemShut {NoStop}%
\bibitem [{\citenamefont {Bouwmeester}\ \emph {et~al.}(1997)\citenamefont
  {Bouwmeester}, \citenamefont {Pan}, \citenamefont {Mattle}, \citenamefont
  {Eibl}, \citenamefont {Weinfurter},\ and\ \citenamefont
  {Zeilinger}}]{Bouwmeester97}%
  \BibitemOpen
  \bibfield  {author} {\bibinfo {author} {\bibfnamefont {D.}~\bibnamefont
  {Bouwmeester}}, \bibinfo {author} {\bibfnamefont {J.-W.}\ \bibnamefont
  {Pan}}, \bibinfo {author} {\bibfnamefont {K.}~\bibnamefont {Mattle}},
  \bibinfo {author} {\bibfnamefont {M.}~\bibnamefont {Eibl}}, \bibinfo {author}
  {\bibfnamefont {H.}~\bibnamefont {Weinfurter}}, \ and\ \bibinfo {author}
  {\bibfnamefont {A.}~\bibnamefont {Zeilinger}},\ }\href@noop {} {\bibfield
  {journal} {\bibinfo  {journal} {Nature}\ }\textbf {\bibinfo {volume} {390}},\
  \bibinfo {pages} {575} (\bibinfo {year} {1997})}\BibitemShut {NoStop}%
\bibitem [{\citenamefont {Yin}\ \emph {et~al.}(2012)\citenamefont {Yin},
  \citenamefont {Ren}, \citenamefont {Lu}, \citenamefont {Cao}, \citenamefont
  {Yong}, \citenamefont {Wu}, \citenamefont {Liu}, \citenamefont {Liao},
  \citenamefont {Zhou}, \citenamefont {Jiang}, \citenamefont {Cai},
  \citenamefont {Xu}, \citenamefont {Pan}, \citenamefont {Jia}, \citenamefont
  {Huang}, \citenamefont {Yin}, \citenamefont {Wang}, \citenamefont {Chen},
  \citenamefont {Peng},\ and\ \citenamefont {Pan}}]{Yin12}%
  \BibitemOpen
  \bibfield  {author} {\bibinfo {author} {\bibfnamefont {J.}~\bibnamefont
  {Yin}}, \bibinfo {author} {\bibfnamefont {J.-G.}\ \bibnamefont {Ren}},
  \bibinfo {author} {\bibfnamefont {H.}~\bibnamefont {Lu}}, \bibinfo {author}
  {\bibfnamefont {Y.}~\bibnamefont {Cao}}, \bibinfo {author} {\bibfnamefont
  {H.-L.}\ \bibnamefont {Yong}}, \bibinfo {author} {\bibfnamefont {Y.-P.}\
  \bibnamefont {Wu}}, \bibinfo {author} {\bibfnamefont {C.}~\bibnamefont
  {Liu}}, \bibinfo {author} {\bibfnamefont {S.-K.}\ \bibnamefont {Liao}},
  \bibinfo {author} {\bibfnamefont {F.}~\bibnamefont {Zhou}}, \bibinfo {author}
  {\bibfnamefont {Y.}~\bibnamefont {Jiang}}, \bibinfo {author} {\bibfnamefont
  {X.-D.}\ \bibnamefont {Cai}}, \bibinfo {author} {\bibfnamefont
  {P.}~\bibnamefont {Xu}}, \bibinfo {author} {\bibfnamefont {G.-S.}\
  \bibnamefont {Pan}}, \bibinfo {author} {\bibfnamefont {J.-J.}\ \bibnamefont
  {Jia}}, \bibinfo {author} {\bibfnamefont {Y.-M.}\ \bibnamefont {Huang}},
  \bibinfo {author} {\bibfnamefont {H.}~\bibnamefont {Yin}}, \bibinfo {author}
  {\bibfnamefont {J.-Y.}\ \bibnamefont {Wang}}, \bibinfo {author}
  {\bibfnamefont {Y.-A.}\ \bibnamefont {Chen}}, \bibinfo {author}
  {\bibfnamefont {C.-Z.}\ \bibnamefont {Peng}}, \ and\ \bibinfo {author}
  {\bibfnamefont {J.-W.}\ \bibnamefont {Pan}},\ }\href@noop {} {\bibfield
  {journal} {\bibinfo  {journal} {Nature}\ }\textbf {\bibinfo {volume} {488}},\
  \bibinfo {pages} {185} (\bibinfo {year} {2012})}\BibitemShut {NoStop}%
\bibitem [{\citenamefont {Ma}\ \emph {et~al.}(2012)\citenamefont {Ma},
  \citenamefont {Herbst}, \citenamefont {Scheidl}, \citenamefont {Wang},
  \citenamefont {Kropatschek}, \citenamefont {Naylor}, \citenamefont
  {Wittmann}, \citenamefont {Mech}, \citenamefont {Kofler}, \citenamefont
  {Anisimova}, \citenamefont {Makarov}, \citenamefont {Jennewein},
  \citenamefont {Ursin},\ and\ \citenamefont {Zeilinger}}]{Ma12}%
  \BibitemOpen
  \bibfield  {author} {\bibinfo {author} {\bibfnamefont {X.-S.}\ \bibnamefont
  {Ma}}, \bibinfo {author} {\bibfnamefont {T.}~\bibnamefont {Herbst}}, \bibinfo
  {author} {\bibfnamefont {T.}~\bibnamefont {Scheidl}}, \bibinfo {author}
  {\bibfnamefont {D.}~\bibnamefont {Wang}}, \bibinfo {author} {\bibfnamefont
  {S.}~\bibnamefont {Kropatschek}}, \bibinfo {author} {\bibfnamefont
  {W.}~\bibnamefont {Naylor}}, \bibinfo {author} {\bibfnamefont
  {B.}~\bibnamefont {Wittmann}}, \bibinfo {author} {\bibfnamefont
  {A.}~\bibnamefont {Mech}}, \bibinfo {author} {\bibfnamefont {J.}~\bibnamefont
  {Kofler}}, \bibinfo {author} {\bibfnamefont {E.}~\bibnamefont {Anisimova}},
  \bibinfo {author} {\bibfnamefont {V.}~\bibnamefont {Makarov}}, \bibinfo
  {author} {\bibfnamefont {T.}~\bibnamefont {Jennewein}}, \bibinfo {author}
  {\bibfnamefont {R.}~\bibnamefont {Ursin}}, \ and\ \bibinfo {author}
  {\bibfnamefont {A.}~\bibnamefont {Zeilinger}},\ }\href@noop {} {\bibfield
  {journal} {\bibinfo  {journal} {Nature}\ }\textbf {\bibinfo {volume} {489}},\
  \bibinfo {pages} {269} (\bibinfo {year} {2012})}\BibitemShut {NoStop}%
\bibitem [{\citenamefont {Lu}\ \emph {et~al.}(2014)\citenamefont {Lu},
  \citenamefont {Chen}, \citenamefont {Liu}, \citenamefont {Xu}, \citenamefont
  {Yao}, \citenamefont {Li}, \citenamefont {Liu}, \citenamefont {Zhao},
  \citenamefont {Chen},\ and\ \citenamefont {Pan}}]{Luhe14}%
  \BibitemOpen
  \bibfield  {author} {\bibinfo {author} {\bibfnamefont {H.}~\bibnamefont
  {Lu}}, \bibinfo {author} {\bibfnamefont {L.-K.}\ \bibnamefont {Chen}},
  \bibinfo {author} {\bibfnamefont {C.}~\bibnamefont {Liu}}, \bibinfo {author}
  {\bibfnamefont {P.}~\bibnamefont {Xu}}, \bibinfo {author} {\bibfnamefont
  {X.-C.}\ \bibnamefont {Yao}}, \bibinfo {author} {\bibfnamefont
  {L.}~\bibnamefont {Li}}, \bibinfo {author} {\bibfnamefont {N.-L.}\
  \bibnamefont {Liu}}, \bibinfo {author} {\bibfnamefont {B.}~\bibnamefont
  {Zhao}}, \bibinfo {author} {\bibfnamefont {Y.-A.}\ \bibnamefont {Chen}}, \
  and\ \bibinfo {author} {\bibfnamefont {J.-W.}\ \bibnamefont {Pan}},\
  }\href@noop {} {\bibfield  {journal} {\bibinfo  {journal} {Nat. Photon.}\
  }\textbf {\bibinfo {volume} {8}},\ \bibinfo {pages} {364} (\bibinfo {year}
  {2014})}\BibitemShut {NoStop}%
\bibitem [{\citenamefont {Lu}\ \emph {et~al.}(2016)\citenamefont {Lu},
  \citenamefont {Zhang}, \citenamefont {Chen}, \citenamefont {Li},
  \citenamefont {Liu}, \citenamefont {Li}, \citenamefont {Liu}, \citenamefont
  {Ma}, \citenamefont {Chen},\ and\ \citenamefont {Pan}}]{Lu16QSS}%
  \BibitemOpen
  \bibfield  {author} {\bibinfo {author} {\bibfnamefont {H.}~\bibnamefont
  {Lu}}, \bibinfo {author} {\bibfnamefont {Z.}~\bibnamefont {Zhang}}, \bibinfo
  {author} {\bibfnamefont {L.-K.}\ \bibnamefont {Chen}}, \bibinfo {author}
  {\bibfnamefont {Z.-D.}\ \bibnamefont {Li}}, \bibinfo {author} {\bibfnamefont
  {C.}~\bibnamefont {Liu}}, \bibinfo {author} {\bibfnamefont {L.}~\bibnamefont
  {Li}}, \bibinfo {author} {\bibfnamefont {N.-L.}\ \bibnamefont {Liu}},
  \bibinfo {author} {\bibfnamefont {X.}~\bibnamefont {Ma}}, \bibinfo {author}
  {\bibfnamefont {Y.-A.}\ \bibnamefont {Chen}}, \ and\ \bibinfo {author}
  {\bibfnamefont {J.-W.}\ \bibnamefont {Pan}},\ }\href@noop {} {\bibfield
  {journal} {\bibinfo  {journal} {Phys. Rev. Lett.}\ }\textbf {\bibinfo
  {volume} {117}},\ \bibinfo {pages} {030501} (\bibinfo {year}
  {2016})}\BibitemShut {NoStop}%
\bibitem [{\citenamefont {Giovannetti}\ \emph {et~al.}(2004)\citenamefont
  {Giovannetti}, \citenamefont {Lloyd},\ and\ \citenamefont
  {Maccone}}]{Giovannetti04}%
  \BibitemOpen
  \bibfield  {author} {\bibinfo {author} {\bibfnamefont {V.}~\bibnamefont
  {Giovannetti}}, \bibinfo {author} {\bibfnamefont {S.}~\bibnamefont {Lloyd}},
  \ and\ \bibinfo {author} {\bibfnamefont {L.}~\bibnamefont {Maccone}},\
  }\href@noop {} {\bibfield  {journal} {\bibinfo  {journal} {Science}\ }\textbf
  {\bibinfo {volume} {306}},\ \bibinfo {pages} {1330} (\bibinfo {year}
  {2004})}\BibitemShut {NoStop}%
\bibitem [{\citenamefont {Knill}\ \emph {et~al.}(2001)\citenamefont {Knill},
  \citenamefont {Laflamme},\ and\ \citenamefont {Milburn}}]{KLM01}%
  \BibitemOpen
  \bibfield  {author} {\bibinfo {author} {\bibfnamefont {E.}~\bibnamefont
  {Knill}}, \bibinfo {author} {\bibfnamefont {R.}~\bibnamefont {Laflamme}}, \
  and\ \bibinfo {author} {\bibfnamefont {G.~J.}\ \bibnamefont {Milburn}},\
  }\href@noop {} {\bibfield  {journal} {\bibinfo  {journal} {Nature}\ }\textbf
  {\bibinfo {volume} {409}},\ \bibinfo {pages} {46} (\bibinfo {year}
  {2001})}\BibitemShut {NoStop}%
\bibitem [{\citenamefont {Kok}\ \emph {et~al.}(2007)\citenamefont {Kok},
  \citenamefont {Munro}, \citenamefont {Nemoto}, \citenamefont {Ralph},
  \citenamefont {Dowling},\ and\ \citenamefont {Milburn}}]{Kok07}%
  \BibitemOpen
  \bibfield  {author} {\bibinfo {author} {\bibfnamefont {P.}~\bibnamefont
  {Kok}}, \bibinfo {author} {\bibfnamefont {W.~J.}\ \bibnamefont {Munro}},
  \bibinfo {author} {\bibfnamefont {K.}~\bibnamefont {Nemoto}}, \bibinfo
  {author} {\bibfnamefont {T.~C.}\ \bibnamefont {Ralph}}, \bibinfo {author}
  {\bibfnamefont {J.~P.}\ \bibnamefont {Dowling}}, \ and\ \bibinfo {author}
  {\bibfnamefont {G.~J.}\ \bibnamefont {Milburn}},\ }\href@noop {} {\bibfield
  {journal} {\bibinfo  {journal} {Rev. Mod. Phys.}\ }\textbf {\bibinfo {volume}
  {79}},\ \bibinfo {pages} {135} (\bibinfo {year} {2007})}\BibitemShut
  {NoStop}%
\bibitem [{\citenamefont {Walther}\ \emph {et~al.}(2005)\citenamefont
  {Walther}, \citenamefont {Resch}, \citenamefont {Rudolph}, \citenamefont
  {Schenck}, \citenamefont {Weinfurter}, \citenamefont {Vedral}, \citenamefont
  {Aspelmeyer},\ and\ \citenamefont {Zeilinger}}]{Walther05}%
  \BibitemOpen
  \bibfield  {author} {\bibinfo {author} {\bibfnamefont {P.}~\bibnamefont
  {Walther}}, \bibinfo {author} {\bibfnamefont {K.~J.}\ \bibnamefont {Resch}},
  \bibinfo {author} {\bibfnamefont {T.}~\bibnamefont {Rudolph}}, \bibinfo
  {author} {\bibfnamefont {E.}~\bibnamefont {Schenck}}, \bibinfo {author}
  {\bibfnamefont {H.}~\bibnamefont {Weinfurter}}, \bibinfo {author}
  {\bibfnamefont {V.}~\bibnamefont {Vedral}}, \bibinfo {author} {\bibfnamefont
  {M.}~\bibnamefont {Aspelmeyer}}, \ and\ \bibinfo {author} {\bibfnamefont
  {A.}~\bibnamefont {Zeilinger}},\ }\href@noop {} {\bibfield  {journal}
  {\bibinfo  {journal} {Nature}\ }\textbf {\bibinfo {volume} {434}},\ \bibinfo
  {pages} {169} (\bibinfo {year} {2005})}\BibitemShut {NoStop}%
\bibitem [{\citenamefont {Chen}\ \emph {et~al.}(2007)\citenamefont {Chen},
  \citenamefont {Li}, \citenamefont {Zhang}, \citenamefont {Chen},
  \citenamefont {Goebel}, \citenamefont {Chen}, \citenamefont {Mair},\ and\
  \citenamefont {Pan}}]{Kai07}%
  \BibitemOpen
  \bibfield  {author} {\bibinfo {author} {\bibfnamefont {K.}~\bibnamefont
  {Chen}}, \bibinfo {author} {\bibfnamefont {C.-M.}\ \bibnamefont {Li}},
  \bibinfo {author} {\bibfnamefont {Q.}~\bibnamefont {Zhang}}, \bibinfo
  {author} {\bibfnamefont {Y.-A.}\ \bibnamefont {Chen}}, \bibinfo {author}
  {\bibfnamefont {A.}~\bibnamefont {Goebel}}, \bibinfo {author} {\bibfnamefont
  {S.}~\bibnamefont {Chen}}, \bibinfo {author} {\bibfnamefont {A.}~\bibnamefont
  {Mair}}, \ and\ \bibinfo {author} {\bibfnamefont {J.-W.}\ \bibnamefont
  {Pan}},\ }\href@noop {} {\bibfield  {journal} {\bibinfo  {journal} {Phys.
  Rev. Lett.}\ }\textbf {\bibinfo {volume} {99}},\ \bibinfo {pages} {120503}
  (\bibinfo {year} {2007})}\BibitemShut {NoStop}%
\bibitem [{\citenamefont {Lu}\ \emph {et~al.}(2007{\natexlab{a}})\citenamefont
  {Lu}, \citenamefont {Browne}, \citenamefont {Yang},\ and\ \citenamefont
  {Pan}}]{Lu07shor}%
  \BibitemOpen
  \bibfield  {author} {\bibinfo {author} {\bibfnamefont {C.-Y.}\ \bibnamefont
  {Lu}}, \bibinfo {author} {\bibfnamefont {D.~E.}\ \bibnamefont {Browne}},
  \bibinfo {author} {\bibfnamefont {T.}~\bibnamefont {Yang}}, \ and\ \bibinfo
  {author} {\bibfnamefont {J.-W.}\ \bibnamefont {Pan}},\ }\href@noop {}
  {\bibfield  {journal} {\bibinfo  {journal} {Phys. Rev. Lett.}\ }\textbf
  {\bibinfo {volume} {99}},\ \bibinfo {pages} {250504} (\bibinfo {year}
  {2007}{\natexlab{a}})}\BibitemShut {NoStop}%
\bibitem [{\citenamefont {Lanyon}\ \emph {et~al.}(2007)\citenamefont {Lanyon},
  \citenamefont {Weinhold}, \citenamefont {Langford}, \citenamefont {Barbieri},
  \citenamefont {James}, \citenamefont {Gilchrist},\ and\ \citenamefont
  {White}}]{Lanyon07}%
  \BibitemOpen
  \bibfield  {author} {\bibinfo {author} {\bibfnamefont {B.~P.}\ \bibnamefont
  {Lanyon}}, \bibinfo {author} {\bibfnamefont {T.~J.}\ \bibnamefont
  {Weinhold}}, \bibinfo {author} {\bibfnamefont {N.~K.}\ \bibnamefont
  {Langford}}, \bibinfo {author} {\bibfnamefont {M.}~\bibnamefont {Barbieri}},
  \bibinfo {author} {\bibfnamefont {D.~F.~V.}\ \bibnamefont {James}}, \bibinfo
  {author} {\bibfnamefont {A.}~\bibnamefont {Gilchrist}}, \ and\ \bibinfo
  {author} {\bibfnamefont {A.~G.}\ \bibnamefont {White}},\ }\href@noop {}
  {\bibfield  {journal} {\bibinfo  {journal} {Phys. Rev. Lett.}\ }\textbf
  {\bibinfo {volume} {99}},\ \bibinfo {pages} {250505} (\bibinfo {year}
  {2007})}\BibitemShut {NoStop}%
\bibitem [{\citenamefont {Yao}\ \emph {et~al.}(2012{\natexlab{a}})\citenamefont
  {Yao}, \citenamefont {Wang}, \citenamefont {Chen}, \citenamefont {Gao},
  \citenamefont {Fowler}, \citenamefont {Raussendorf}, \citenamefont {Chen},
  \citenamefont {Liu}, \citenamefont {Lu}, \citenamefont {Deng}, \citenamefont
  {Chen},\ and\ \citenamefont {Pan}}]{Yao12TEC}%
  \BibitemOpen
  \bibfield  {author} {\bibinfo {author} {\bibfnamefont {X.-C.}\ \bibnamefont
  {Yao}}, \bibinfo {author} {\bibfnamefont {T.-X.}\ \bibnamefont {Wang}},
  \bibinfo {author} {\bibfnamefont {H.-Z.}\ \bibnamefont {Chen}}, \bibinfo
  {author} {\bibfnamefont {W.-B.}\ \bibnamefont {Gao}}, \bibinfo {author}
  {\bibfnamefont {A.~G.}\ \bibnamefont {Fowler}}, \bibinfo {author}
  {\bibfnamefont {R.}~\bibnamefont {Raussendorf}}, \bibinfo {author}
  {\bibfnamefont {Z.-B.}\ \bibnamefont {Chen}}, \bibinfo {author}
  {\bibfnamefont {N.-L.}\ \bibnamefont {Liu}}, \bibinfo {author} {\bibfnamefont
  {C.-Y.}\ \bibnamefont {Lu}}, \bibinfo {author} {\bibfnamefont {Y.-J.}\
  \bibnamefont {Deng}}, \bibinfo {author} {\bibfnamefont {Y.-A.}\ \bibnamefont
  {Chen}}, \ and\ \bibinfo {author} {\bibfnamefont {J.-W.}\ \bibnamefont
  {Pan}},\ }\href@noop {} {\bibfield  {journal} {\bibinfo  {journal} {Nature}\
  }\textbf {\bibinfo {volume} {482}},\ \bibinfo {pages} {489} (\bibinfo {year}
  {2012}{\natexlab{a}})}\BibitemShut {NoStop}%
\bibitem [{\citenamefont {Cai}\ \emph {et~al.}(2013)\citenamefont {Cai},
  \citenamefont {Weedbrook}, \citenamefont {Su}, \citenamefont {Chen},
  \citenamefont {Gu}, \citenamefont {Zhu}, \citenamefont {Li}, \citenamefont
  {Liu}, \citenamefont {Lu},\ and\ \citenamefont {Pan}}]{Cai13}%
  \BibitemOpen
  \bibfield  {author} {\bibinfo {author} {\bibfnamefont {X.-D.}\ \bibnamefont
  {Cai}}, \bibinfo {author} {\bibfnamefont {C.}~\bibnamefont {Weedbrook}},
  \bibinfo {author} {\bibfnamefont {Z.-E.}\ \bibnamefont {Su}}, \bibinfo
  {author} {\bibfnamefont {M.-C.}\ \bibnamefont {Chen}}, \bibinfo {author}
  {\bibfnamefont {M.}~\bibnamefont {Gu}}, \bibinfo {author} {\bibfnamefont
  {M.-J.}\ \bibnamefont {Zhu}}, \bibinfo {author} {\bibfnamefont
  {L.}~\bibnamefont {Li}}, \bibinfo {author} {\bibfnamefont {N.-L.}\
  \bibnamefont {Liu}}, \bibinfo {author} {\bibfnamefont {C.-Y.}\ \bibnamefont
  {Lu}}, \ and\ \bibinfo {author} {\bibfnamefont {J.-W.}\ \bibnamefont {Pan}},\
  }\href@noop {} {\bibfield  {journal} {\bibinfo  {journal} {Phys. Rev. Lett.}\
  }\textbf {\bibinfo {volume} {110}},\ \bibinfo {pages} {230501} (\bibinfo
  {year} {2013})}\BibitemShut {NoStop}%
\bibitem [{\citenamefont {Spring}\ \emph {et~al.}(2013)\citenamefont {Spring},
  \citenamefont {Metcalf}, \citenamefont {Humphreys}, \citenamefont
  {Kolthammer}, \citenamefont {Jin}, \citenamefont {Barbieri}, \citenamefont
  {Datta}, \citenamefont {Thomas-Peter}, \citenamefont {Langford},
  \citenamefont {Kundys}, \citenamefont {Gates}, \citenamefont {Smith},
  \citenamefont {Smith},\ and\ \citenamefont {Walmsley}}]{Spring13}%
  \BibitemOpen
  \bibfield  {author} {\bibinfo {author} {\bibfnamefont {J.~B.}\ \bibnamefont
  {Spring}}, \bibinfo {author} {\bibfnamefont {B.~J.}\ \bibnamefont {Metcalf}},
  \bibinfo {author} {\bibfnamefont {P.~C.}\ \bibnamefont {Humphreys}}, \bibinfo
  {author} {\bibfnamefont {W.~S.}\ \bibnamefont {Kolthammer}}, \bibinfo
  {author} {\bibfnamefont {X.-M.}\ \bibnamefont {Jin}}, \bibinfo {author}
  {\bibfnamefont {M.}~\bibnamefont {Barbieri}}, \bibinfo {author}
  {\bibfnamefont {A.}~\bibnamefont {Datta}}, \bibinfo {author} {\bibfnamefont
  {N.}~\bibnamefont {Thomas-Peter}}, \bibinfo {author} {\bibfnamefont {N.~K.}\
  \bibnamefont {Langford}}, \bibinfo {author} {\bibfnamefont {D.}~\bibnamefont
  {Kundys}}, \bibinfo {author} {\bibfnamefont {J.~C.}\ \bibnamefont {Gates}},
  \bibinfo {author} {\bibfnamefont {B.~J.}\ \bibnamefont {Smith}}, \bibinfo
  {author} {\bibfnamefont {P.~G.~R.}\ \bibnamefont {Smith}}, \ and\ \bibinfo
  {author} {\bibfnamefont {I.~A.}\ \bibnamefont {Walmsley}},\ }\href@noop {}
  {\bibfield  {journal} {\bibinfo  {journal} {Science}\ }\textbf {\bibinfo
  {volume} {339}},\ \bibinfo {pages} {798} (\bibinfo {year}
  {2013})}\BibitemShut {NoStop}%
\bibitem [{\citenamefont {Broome}\ \emph {et~al.}(2013)\citenamefont {Broome},
  \citenamefont {Fedrizzi}, \citenamefont {Rahimi-Keshari}, \citenamefont
  {Dove}, \citenamefont {Aaronson}, \citenamefont {Ralph},\ and\ \citenamefont
  {White}}]{Broome13}%
  \BibitemOpen
  \bibfield  {author} {\bibinfo {author} {\bibfnamefont {M.~A.}\ \bibnamefont
  {Broome}}, \bibinfo {author} {\bibfnamefont {A.}~\bibnamefont {Fedrizzi}},
  \bibinfo {author} {\bibfnamefont {S.}~\bibnamefont {Rahimi-Keshari}},
  \bibinfo {author} {\bibfnamefont {J.}~\bibnamefont {Dove}}, \bibinfo {author}
  {\bibfnamefont {S.}~\bibnamefont {Aaronson}}, \bibinfo {author}
  {\bibfnamefont {T.~C.}\ \bibnamefont {Ralph}}, \ and\ \bibinfo {author}
  {\bibfnamefont {A.~G.}\ \bibnamefont {White}},\ }\href@noop {} {\bibfield
  {journal} {\bibinfo  {journal} {Science}\ }\textbf {\bibinfo {volume}
  {339}},\ \bibinfo {pages} {794} (\bibinfo {year} {2013})}\BibitemShut
  {NoStop}%
\bibitem [{\citenamefont {Tillmann}\ \emph {et~al.}(2013)\citenamefont
  {Tillmann}, \citenamefont {Daki{\'c}}, \citenamefont {Heilmann},
  \citenamefont {Nolte}, \citenamefont {Szameit},\ and\ \citenamefont
  {Walther}}]{Tillmann13}%
  \BibitemOpen
  \bibfield  {author} {\bibinfo {author} {\bibfnamefont {M.}~\bibnamefont
  {Tillmann}}, \bibinfo {author} {\bibfnamefont {B.}~\bibnamefont {Daki{\'c}}},
  \bibinfo {author} {\bibfnamefont {R.}~\bibnamefont {Heilmann}}, \bibinfo
  {author} {\bibfnamefont {S.}~\bibnamefont {Nolte}}, \bibinfo {author}
  {\bibfnamefont {A.}~\bibnamefont {Szameit}}, \ and\ \bibinfo {author}
  {\bibfnamefont {P.}~\bibnamefont {Walther}},\ }\href@noop {} {\bibfield
  {journal} {\bibinfo  {journal} {Nat. Photon.}\ }\textbf {\bibinfo {volume}
  {7}},\ \bibinfo {pages} {540} (\bibinfo {year} {2013})}\BibitemShut {NoStop}%
\bibitem [{\citenamefont {Crespi}\ \emph {et~al.}(2013)\citenamefont {Crespi},
  \citenamefont {Osellame}, \citenamefont {Ramponi}, \citenamefont {Brod},
  \citenamefont {Galv{\~a}o}, \citenamefont {Spagnolo}, \citenamefont
  {Vitelli}, \citenamefont {Maiorino}, \citenamefont {Mataloni},\ and\
  \citenamefont {Sciarrino}}]{Crespi13}%
  \BibitemOpen
  \bibfield  {author} {\bibinfo {author} {\bibfnamefont {A.}~\bibnamefont
  {Crespi}}, \bibinfo {author} {\bibfnamefont {R.}~\bibnamefont {Osellame}},
  \bibinfo {author} {\bibfnamefont {R.}~\bibnamefont {Ramponi}}, \bibinfo
  {author} {\bibfnamefont {D.~J.}\ \bibnamefont {Brod}}, \bibinfo {author}
  {\bibfnamefont {E.~F.}\ \bibnamefont {Galv{\~a}o}}, \bibinfo {author}
  {\bibfnamefont {N.}~\bibnamefont {Spagnolo}}, \bibinfo {author}
  {\bibfnamefont {C.}~\bibnamefont {Vitelli}}, \bibinfo {author} {\bibfnamefont
  {E.}~\bibnamefont {Maiorino}}, \bibinfo {author} {\bibfnamefont
  {P.}~\bibnamefont {Mataloni}}, \ and\ \bibinfo {author} {\bibfnamefont
  {F.}~\bibnamefont {Sciarrino}},\ }\href@noop {} {\bibfield  {journal}
  {\bibinfo  {journal} {Nat. Photon.}\ }\textbf {\bibinfo {volume} {7}},\
  \bibinfo {pages} {545} (\bibinfo {year} {2013})}\BibitemShut {NoStop}%
\bibitem [{\citenamefont {Spagnolo}\ \emph {et~al.}(2014)\citenamefont
  {Spagnolo}, \citenamefont {Vitelli}, \citenamefont {Bentivegna},
  \citenamefont {Brod}, \citenamefont {Crespi}, \citenamefont {Flamini},
  \citenamefont {Giacomini}, \citenamefont {Milani}, \citenamefont {Ramponi},
  \citenamefont {Mataloni}, \citenamefont {Osellame}, \citenamefont {Galvao},\
  and\ \citenamefont {Sciarrino}}]{Spagnolo14}%
  \BibitemOpen
  \bibfield  {author} {\bibinfo {author} {\bibfnamefont {N.}~\bibnamefont
  {Spagnolo}}, \bibinfo {author} {\bibfnamefont {C.}~\bibnamefont {Vitelli}},
  \bibinfo {author} {\bibfnamefont {M.}~\bibnamefont {Bentivegna}}, \bibinfo
  {author} {\bibfnamefont {D.~J.}\ \bibnamefont {Brod}}, \bibinfo {author}
  {\bibfnamefont {A.}~\bibnamefont {Crespi}}, \bibinfo {author} {\bibfnamefont
  {F.}~\bibnamefont {Flamini}}, \bibinfo {author} {\bibfnamefont
  {S.}~\bibnamefont {Giacomini}}, \bibinfo {author} {\bibfnamefont
  {G.}~\bibnamefont {Milani}}, \bibinfo {author} {\bibfnamefont
  {R.}~\bibnamefont {Ramponi}}, \bibinfo {author} {\bibfnamefont
  {P.}~\bibnamefont {Mataloni}}, \bibinfo {author} {\bibfnamefont
  {R.}~\bibnamefont {Osellame}}, \bibinfo {author} {\bibfnamefont {E.~F.}\
  \bibnamefont {Galvao}}, \ and\ \bibinfo {author} {\bibfnamefont
  {F.}~\bibnamefont {Sciarrino}},\ }\href@noop {} {\bibfield  {journal}
  {\bibinfo  {journal} {Nat. Photon.}\ }\textbf {\bibinfo {volume} {8}},\
  \bibinfo {pages} {615} (\bibinfo {year} {2014})}\BibitemShut {NoStop}%
\bibitem [{\citenamefont {Carolan}\ \emph {et~al.}(2015)\citenamefont
  {Carolan}, \citenamefont {Harrold}, \citenamefont {Sparrow}, \citenamefont
  {Mart{\'\i}n-L{\'o}pez}, \citenamefont {Russell}, \citenamefont
  {Silverstone}, \citenamefont {Shadbolt}, \citenamefont {Matsuda},
  \citenamefont {Oguma}, \citenamefont {Itoh}, \citenamefont {Marshall},
  \citenamefont {Thompson}, \citenamefont {Matthews}, \citenamefont
  {Hashimoto}, \citenamefont {O{\textquoteright}Brien},\ and\ \citenamefont
  {Laing}}]{Carolan15}%
  \BibitemOpen
  \bibfield  {author} {\bibinfo {author} {\bibfnamefont {J.}~\bibnamefont
  {Carolan}}, \bibinfo {author} {\bibfnamefont {C.}~\bibnamefont {Harrold}},
  \bibinfo {author} {\bibfnamefont {C.}~\bibnamefont {Sparrow}}, \bibinfo
  {author} {\bibfnamefont {E.}~\bibnamefont {Mart{\'\i}n-L{\'o}pez}}, \bibinfo
  {author} {\bibfnamefont {N.~J.}\ \bibnamefont {Russell}}, \bibinfo {author}
  {\bibfnamefont {J.~W.}\ \bibnamefont {Silverstone}}, \bibinfo {author}
  {\bibfnamefont {P.~J.}\ \bibnamefont {Shadbolt}}, \bibinfo {author}
  {\bibfnamefont {N.}~\bibnamefont {Matsuda}}, \bibinfo {author} {\bibfnamefont
  {M.}~\bibnamefont {Oguma}}, \bibinfo {author} {\bibfnamefont
  {M.}~\bibnamefont {Itoh}}, \bibinfo {author} {\bibfnamefont {G.~D.}\
  \bibnamefont {Marshall}}, \bibinfo {author} {\bibfnamefont {M.~G.}\
  \bibnamefont {Thompson}}, \bibinfo {author} {\bibfnamefont {J.~C.~F.}\
  \bibnamefont {Matthews}}, \bibinfo {author} {\bibfnamefont {T.}~\bibnamefont
  {Hashimoto}}, \bibinfo {author} {\bibfnamefont {J.~L.}\ \bibnamefont
  {O{\textquoteright}Brien}}, \ and\ \bibinfo {author} {\bibfnamefont
  {A.}~\bibnamefont {Laing}},\ }\href@noop {} {\bibfield  {journal} {\bibinfo
  {journal} {Science}\ }\textbf {\bibinfo {volume} {349}},\ \bibinfo {pages}
  {711} (\bibinfo {year} {2015})}\BibitemShut {NoStop}%
\bibitem [{\citenamefont {Wang}\ \emph {et~al.}(2015)\citenamefont {Wang},
  \citenamefont {Cai}, \citenamefont {Su}, \citenamefont {Chen}, \citenamefont
  {Wu}, \citenamefont {Li}, \citenamefont {Liu}, \citenamefont {Lu},\ and\
  \citenamefont {Pan}}]{Wang15}%
  \BibitemOpen
  \bibfield  {author} {\bibinfo {author} {\bibfnamefont {X.-L.}\ \bibnamefont
  {Wang}}, \bibinfo {author} {\bibfnamefont {X.-D.}\ \bibnamefont {Cai}},
  \bibinfo {author} {\bibfnamefont {Z.-E.}\ \bibnamefont {Su}}, \bibinfo
  {author} {\bibfnamefont {M.-C.}\ \bibnamefont {Chen}}, \bibinfo {author}
  {\bibfnamefont {D.}~\bibnamefont {Wu}}, \bibinfo {author} {\bibfnamefont
  {L.}~\bibnamefont {Li}}, \bibinfo {author} {\bibfnamefont {N.-L.}\
  \bibnamefont {Liu}}, \bibinfo {author} {\bibfnamefont {C.-Y.}\ \bibnamefont
  {Lu}}, \ and\ \bibinfo {author} {\bibfnamefont {J.-W.}\ \bibnamefont {Pan}},\
  }\href@noop {} {\bibfield  {journal} {\bibinfo  {journal} {Nature}\ }\textbf
  {\bibinfo {volume} {518}},\ \bibinfo {pages} {516} (\bibinfo {year}
  {2015})}\BibitemShut {NoStop}%
\bibitem [{\citenamefont {Bouwmeester}\ \emph {et~al.}(1999)\citenamefont
  {Bouwmeester}, \citenamefont {Pan}, \citenamefont {Daniell}, \citenamefont
  {Weinfurter},\ and\ \citenamefont {Zeilinger}}]{Bouwmeester99}%
  \BibitemOpen
  \bibfield  {author} {\bibinfo {author} {\bibfnamefont {D.}~\bibnamefont
  {Bouwmeester}}, \bibinfo {author} {\bibfnamefont {J.-W.}\ \bibnamefont
  {Pan}}, \bibinfo {author} {\bibfnamefont {M.}~\bibnamefont {Daniell}},
  \bibinfo {author} {\bibfnamefont {H.}~\bibnamefont {Weinfurter}}, \ and\
  \bibinfo {author} {\bibfnamefont {A.}~\bibnamefont {Zeilinger}},\ }\href@noop
  {} {\bibfield  {journal} {\bibinfo  {journal} {Phys. Rev. Lett.}\ }\textbf
  {\bibinfo {volume} {82}},\ \bibinfo {pages} {1345} (\bibinfo {year}
  {1999})}\BibitemShut {NoStop}%
\bibitem [{\citenamefont {Pan}\ \emph {et~al.}(2001)\citenamefont {Pan},
  \citenamefont {Daniell}, \citenamefont {Gasparoni}, \citenamefont {Weihs},\
  and\ \citenamefont {Zeilinger}}]{Pan01prl}%
  \BibitemOpen
  \bibfield  {author} {\bibinfo {author} {\bibfnamefont {J.-W.}\ \bibnamefont
  {Pan}}, \bibinfo {author} {\bibfnamefont {M.}~\bibnamefont {Daniell}},
  \bibinfo {author} {\bibfnamefont {S.}~\bibnamefont {Gasparoni}}, \bibinfo
  {author} {\bibfnamefont {G.}~\bibnamefont {Weihs}}, \ and\ \bibinfo {author}
  {\bibfnamefont {A.}~\bibnamefont {Zeilinger}},\ }\href@noop {} {\bibfield
  {journal} {\bibinfo  {journal} {Phys. Rev. Lett.}\ }\textbf {\bibinfo
  {volume} {86}},\ \bibinfo {pages} {4435} (\bibinfo {year}
  {2001})}\BibitemShut {NoStop}%
\bibitem [{\citenamefont {Zhao}\ \emph {et~al.}(2004)\citenamefont {Zhao},
  \citenamefont {Chen}, \citenamefont {Zhang}, \citenamefont {Yang},
  \citenamefont {Briegel},\ and\ \citenamefont {Pan}}]{Zhao04}%
  \BibitemOpen
  \bibfield  {author} {\bibinfo {author} {\bibfnamefont {Z.}~\bibnamefont
  {Zhao}}, \bibinfo {author} {\bibfnamefont {Y.-A.}\ \bibnamefont {Chen}},
  \bibinfo {author} {\bibfnamefont {A.-N.}\ \bibnamefont {Zhang}}, \bibinfo
  {author} {\bibfnamefont {T.}~\bibnamefont {Yang}}, \bibinfo {author}
  {\bibfnamefont {H.~J.}\ \bibnamefont {Briegel}}, \ and\ \bibinfo {author}
  {\bibfnamefont {J.-W.}\ \bibnamefont {Pan}},\ }\href@noop {} {\bibfield
  {journal} {\bibinfo  {journal} {Nature}\ }\textbf {\bibinfo {volume} {430}},\
  \bibinfo {pages} {54} (\bibinfo {year} {2004})}\BibitemShut {NoStop}%
\bibitem [{\citenamefont {Lu}\ \emph {et~al.}(2007{\natexlab{b}})\citenamefont
  {Lu}, \citenamefont {Zhou}, \citenamefont {G{\"u}hne}, \citenamefont {Gao},
  \citenamefont {Zhang}, \citenamefont {Yuan}, \citenamefont {Goebel},
  \citenamefont {Yang},\ and\ \citenamefont {Pan}}]{Lu07}%
  \BibitemOpen
  \bibfield  {author} {\bibinfo {author} {\bibfnamefont {C.-Y.}\ \bibnamefont
  {Lu}}, \bibinfo {author} {\bibfnamefont {X.-Q.}\ \bibnamefont {Zhou}},
  \bibinfo {author} {\bibfnamefont {O.}~\bibnamefont {G{\"u}hne}}, \bibinfo
  {author} {\bibfnamefont {W.-B.}\ \bibnamefont {Gao}}, \bibinfo {author}
  {\bibfnamefont {J.}~\bibnamefont {Zhang}}, \bibinfo {author} {\bibfnamefont
  {Z.-S.}\ \bibnamefont {Yuan}}, \bibinfo {author} {\bibfnamefont
  {A.}~\bibnamefont {Goebel}}, \bibinfo {author} {\bibfnamefont
  {T.}~\bibnamefont {Yang}}, \ and\ \bibinfo {author} {\bibfnamefont {J.-W.}\
  \bibnamefont {Pan}},\ }\href@noop {} {\bibfield  {journal} {\bibinfo
  {journal} {Nat. Phys.}\ }\textbf {\bibinfo {volume} {3}},\ \bibinfo {pages}
  {91} (\bibinfo {year} {2007}{\natexlab{b}})}\BibitemShut {NoStop}%
\bibitem [{\citenamefont {Yao}\ \emph {et~al.}(2012{\natexlab{b}})\citenamefont
  {Yao}, \citenamefont {Wang}, \citenamefont {Xu}, \citenamefont {Lu},
  \citenamefont {Pan}, \citenamefont {Bao}, \citenamefont {Peng}, \citenamefont
  {Lu}, \citenamefont {Chen},\ and\ \citenamefont {Pan}}]{Yao12}%
  \BibitemOpen
  \bibfield  {author} {\bibinfo {author} {\bibfnamefont {X.-C.}\ \bibnamefont
  {Yao}}, \bibinfo {author} {\bibfnamefont {T.-X.}\ \bibnamefont {Wang}},
  \bibinfo {author} {\bibfnamefont {P.}~\bibnamefont {Xu}}, \bibinfo {author}
  {\bibfnamefont {H.}~\bibnamefont {Lu}}, \bibinfo {author} {\bibfnamefont
  {G.-S.}\ \bibnamefont {Pan}}, \bibinfo {author} {\bibfnamefont {X.-H.}\
  \bibnamefont {Bao}}, \bibinfo {author} {\bibfnamefont {C.-Z.}\ \bibnamefont
  {Peng}}, \bibinfo {author} {\bibfnamefont {C.-Y.}\ \bibnamefont {Lu}},
  \bibinfo {author} {\bibfnamefont {Y.-A.}\ \bibnamefont {Chen}}, \ and\
  \bibinfo {author} {\bibfnamefont {J.-W.}\ \bibnamefont {Pan}},\ }\href@noop
  {} {\bibfield  {journal} {\bibinfo  {journal} {Nat. Photon.}\ }\textbf
  {\bibinfo {volume} {6}},\ \bibinfo {pages} {225} (\bibinfo {year}
  {2012}{\natexlab{b}})}\BibitemShut {NoStop}%
\bibitem [{\citenamefont {Huang}\ \emph {et~al.}(2011)\citenamefont {Huang},
  \citenamefont {Liu}, \citenamefont {Peng}, \citenamefont {Li}, \citenamefont
  {Li}, \citenamefont {Li},\ and\ \citenamefont {Guo}}]{Huang11}%
  \BibitemOpen
  \bibfield  {author} {\bibinfo {author} {\bibfnamefont {Y.-F.}\ \bibnamefont
  {Huang}}, \bibinfo {author} {\bibfnamefont {B.-H.}\ \bibnamefont {Liu}},
  \bibinfo {author} {\bibfnamefont {L.}~\bibnamefont {Peng}}, \bibinfo {author}
  {\bibfnamefont {Y.-H.}\ \bibnamefont {Li}}, \bibinfo {author} {\bibfnamefont
  {L.}~\bibnamefont {Li}}, \bibinfo {author} {\bibfnamefont {C.-F.}\
  \bibnamefont {Li}}, \ and\ \bibinfo {author} {\bibfnamefont {G.-C.}\
  \bibnamefont {Guo}},\ }\href@noop {} {\bibfield  {journal} {\bibinfo
  {journal} {Nat. commun}\ }\textbf {\bibinfo {volume} {2}},\ \bibinfo {pages}
  {546} (\bibinfo {year} {2011})}\BibitemShut {NoStop}%
\bibitem [{\citenamefont {Aaronson}\ and\ \citenamefont
  {Arkhipov}(2011)}]{Aaronson11}%
  \BibitemOpen
  \bibfield  {author} {\bibinfo {author} {\bibfnamefont {S.}~\bibnamefont
  {Aaronson}}\ and\ \bibinfo {author} {\bibfnamefont {A.}~\bibnamefont
  {Arkhipov}},\ }in\ \href@noop {} {\emph {\bibinfo {booktitle} {Proceedings of
  the forty-third annual ACM symposium on Theory of computing}}}\ (\bibinfo
  {organization} {ACM},\ \bibinfo {year} {2011})\ pp.\ \bibinfo {pages}
  {333--342}\BibitemShut {NoStop}%
\bibitem [{\citenamefont {Kwiat}\ \emph {et~al.}(1995)\citenamefont {Kwiat},
  \citenamefont {Mattle}, \citenamefont {Weinfurter}, \citenamefont
  {Zeilinger}, \citenamefont {Sergienko},\ and\ \citenamefont
  {Shih}}]{Kwiat95}%
  \BibitemOpen
  \bibfield  {author} {\bibinfo {author} {\bibfnamefont {P.~G.}\ \bibnamefont
  {Kwiat}}, \bibinfo {author} {\bibfnamefont {K.}~\bibnamefont {Mattle}},
  \bibinfo {author} {\bibfnamefont {H.}~\bibnamefont {Weinfurter}}, \bibinfo
  {author} {\bibfnamefont {A.}~\bibnamefont {Zeilinger}}, \bibinfo {author}
  {\bibfnamefont {A.~V.}\ \bibnamefont {Sergienko}}, \ and\ \bibinfo {author}
  {\bibfnamefont {Y.}~\bibnamefont {Shih}},\ }\href@noop {} {\bibfield
  {journal} {\bibinfo  {journal} {Phys. Rev. Lett.}\ }\textbf {\bibinfo
  {volume} {75}},\ \bibinfo {pages} {4337} (\bibinfo {year}
  {1995})}\BibitemShut {NoStop}%
\bibitem [{not({\natexlab{a}})}]{note1}%
  \BibitemOpen
  \href@noop {} {\  ({\natexlab{a}})},\ \bibinfo {note} {the twofold
  coincidence counting rate of entangled-photon pairs, also known as the
  brightness of entangled-photon pairs, can be described as $R_{T}\xi^{2}$,
  while entanglement decays with $\sim O(1/R_{T})$.}\BibitemShut {Stop}%
\bibitem [{\citenamefont {Laskowski}\ \emph {et~al.}(2009)\citenamefont
  {Laskowski}, \citenamefont {Wie{\'s}niak}, \citenamefont {{\.Z}ukowski},
  \citenamefont {Bourennane},\ and\ \citenamefont {Weinfurter}}]{Laskowski09}%
  \BibitemOpen
  \bibfield  {author} {\bibinfo {author} {\bibfnamefont {W.}~\bibnamefont
  {Laskowski}}, \bibinfo {author} {\bibfnamefont {M.}~\bibnamefont
  {Wie{\'s}niak}}, \bibinfo {author} {\bibfnamefont {M.}~\bibnamefont
  {{\.Z}ukowski}}, \bibinfo {author} {\bibfnamefont {M.}~\bibnamefont
  {Bourennane}}, \ and\ \bibinfo {author} {\bibfnamefont {H.}~\bibnamefont
  {Weinfurter}},\ }\href@noop {} {\bibfield  {journal} {\bibinfo  {journal}
  {Journal of Physics B: Atomic, Molecular and Optical Physics}\ }\textbf
  {\bibinfo {volume} {42}},\ \bibinfo {pages} {114004} (\bibinfo {year}
  {2009})}\BibitemShut {NoStop}%
\bibitem [{not({\natexlab{b}})}]{note7}%
  \BibitemOpen
  \href@noop {} {\  ({\natexlab{b}})},\ \bibinfo {note} {the spatial walk-offs
  result in the decay of beam quality, reducing the SPDC photons' collection
  efficiency.}\BibitemShut {Stop}%
\bibitem [{\citenamefont {Ling}\ \emph {et~al.}(2008)\citenamefont {Ling},
  \citenamefont {Lamas-Linares},\ and\ \citenamefont {Kurtsiefer}}]{Ling08}%
  \BibitemOpen
  \bibfield  {author} {\bibinfo {author} {\bibfnamefont {A.}~\bibnamefont
  {Ling}}, \bibinfo {author} {\bibfnamefont {A.}~\bibnamefont {Lamas-Linares}},
  \ and\ \bibinfo {author} {\bibfnamefont {C.}~\bibnamefont {Kurtsiefer}},\
  }\href@noop {} {\bibfield  {journal} {\bibinfo  {journal} {Phys. Rev.A}\
  }\textbf {\bibinfo {volume} {77}},\ \bibinfo {pages} {043834} (\bibinfo
  {year} {2008})}\BibitemShut {NoStop}%
\bibitem [{not({\natexlab{c}})}]{note2}%
  \BibitemOpen
  \href@noop {} {\  ({\natexlab{c}})},\ \bibinfo {note} {periodically poled
  KTiOPO$_{4}$ (ppKTP) can also meets the two requirements. However, the strong
  frequency correlation in our interested wavelength range prevents ppKTP from
  being a appropriate candidate for the demonstration of multi-photon
  entanglement.}\BibitemShut {Stop}%
\bibitem [{\citenamefont {Rangarajan}\ \emph {et~al.}(2009)\citenamefont
  {Rangarajan}, \citenamefont {Goggin},\ and\ \citenamefont
  {Kwiat}}]{Rangarajan09}%
  \BibitemOpen
  \bibfield  {author} {\bibinfo {author} {\bibfnamefont {R.}~\bibnamefont
  {Rangarajan}}, \bibinfo {author} {\bibfnamefont {M.}~\bibnamefont {Goggin}},
  \ and\ \bibinfo {author} {\bibfnamefont {P.}~\bibnamefont {Kwiat}},\
  }\href@noop {} {\bibfield  {journal} {\bibinfo  {journal} {Opt. express}\
  }\textbf {\bibinfo {volume} {17}},\ \bibinfo {pages} {18920} (\bibinfo {year}
  {2009})}\BibitemShut {NoStop}%
\bibitem [{\citenamefont {Halevy}\ \emph {et~al.}(2011)\citenamefont {Halevy},
  \citenamefont {Megidish}, \citenamefont {Dovrat}, \citenamefont {Eisenberg},
  \citenamefont {Becker},\ and\ \citenamefont {Bohat\'{y}}}]{halevy11}%
  \BibitemOpen
  \bibfield  {author} {\bibinfo {author} {\bibfnamefont {A.}~\bibnamefont
  {Halevy}}, \bibinfo {author} {\bibfnamefont {E.}~\bibnamefont {Megidish}},
  \bibinfo {author} {\bibfnamefont {L.}~\bibnamefont {Dovrat}}, \bibinfo
  {author} {\bibfnamefont {H.}~\bibnamefont {Eisenberg}}, \bibinfo {author}
  {\bibfnamefont {P.}~\bibnamefont {Becker}}, \ and\ \bibinfo {author}
  {\bibfnamefont {L.}~\bibnamefont {Bohat\'{y}}},\ }\href@noop {} {\bibfield
  {journal} {\bibinfo  {journal} {Opt. Express}\ }\textbf {\bibinfo {volume}
  {19}},\ \bibinfo {pages} {20420} (\bibinfo {year} {2011})}\BibitemShut
  {NoStop}%
\bibitem [{\citenamefont {Kim}\ \emph {et~al.}(2003)\citenamefont {Kim},
  \citenamefont {Kulik}, \citenamefont {Chekhova}, \citenamefont {Grice},\ and\
  \citenamefont {Shih}}]{Kim03}%
  \BibitemOpen
  \bibfield  {author} {\bibinfo {author} {\bibfnamefont {Y.-H.}\ \bibnamefont
  {Kim}}, \bibinfo {author} {\bibfnamefont {S.~P.}\ \bibnamefont {Kulik}},
  \bibinfo {author} {\bibfnamefont {M.~V.}\ \bibnamefont {Chekhova}}, \bibinfo
  {author} {\bibfnamefont {W.~P.}\ \bibnamefont {Grice}}, \ and\ \bibinfo
  {author} {\bibfnamefont {Y.}~\bibnamefont {Shih}},\ }\href@noop {} {\bibfield
   {journal} {\bibinfo  {journal} {Phys. Rev. A}\ }\textbf {\bibinfo {volume}
  {67}},\ \bibinfo {pages} {010301} (\bibinfo {year} {2003})}\BibitemShut
  {NoStop}%
\bibitem [{not({\natexlab{d}})}]{note3}%
  \BibitemOpen
  \href@noop {} {\  ({\natexlab{d}})},\ \bibinfo {note} {a similar but not
  identical results is revealed in Ref. \cite{halevy11}.}\BibitemShut {Stop}%
\bibitem [{not({\natexlab{e}})}]{note8}%
  \BibitemOpen
  \href@noop {} {\  ({\natexlab{e}})},\ \bibinfo {note} {the two birefringent
  compensators make the SPDC photons overlap in spatio-temporal mode. However,
  the distortions caused by birefringent walk-off cannot be
  eliminated.}\BibitemShut {Stop}%
\bibitem [{\citenamefont {Grice}\ \emph {et~al.}(2001)\citenamefont {Grice},
  \citenamefont {U'Ren},\ and\ \citenamefont {Walmsley}}]{Grice01}%
  \BibitemOpen
  \bibfield  {author} {\bibinfo {author} {\bibfnamefont {W.~P.}\ \bibnamefont
  {Grice}}, \bibinfo {author} {\bibfnamefont {A.~B.}\ \bibnamefont {U'Ren}}, \
  and\ \bibinfo {author} {\bibfnamefont {I.~A.}\ \bibnamefont {Walmsley}},\
  }\href@noop {} {\bibfield  {journal} {\bibinfo  {journal} {Phys. Rev. A}\
  }\textbf {\bibinfo {volume} {64}},\ \bibinfo {pages} {063815} (\bibinfo
  {year} {2001})}\BibitemShut {NoStop}%
\bibitem [{not({\natexlab{f}})}]{note4}%
  \BibitemOpen
  \href@noop {} {\  ({\natexlab{f}})},\ \bibinfo {note} {the measured ratio of
  the $|HH\rangle$ and $|VV\rangle$ components in each entangled-photon pairs
  is 1.31, 1.29, 1.31, 0.77 and 0.76, respectively when bandpass filters were
  absent. The unbalance even got server when using bandpass filters since
  signal photons in these two components possess different FWHM
  values}\BibitemShut {NoStop}%
\bibitem [{\citenamefont {Hong}\ \emph {et~al.}(1987)\citenamefont {Hong},
  \citenamefont {Ou},\ and\ \citenamefont {Mandel}}]{HOM87}%
  \BibitemOpen
  \bibfield  {author} {\bibinfo {author} {\bibfnamefont {C.~K.}\ \bibnamefont
  {Hong}}, \bibinfo {author} {\bibfnamefont {Z.~Y.}\ \bibnamefont {Ou}}, \ and\
  \bibinfo {author} {\bibfnamefont {L.}~\bibnamefont {Mandel}},\ }\href
  {\doibase 10.1103/PhysRevLett.59.2044} {\bibfield  {journal} {\bibinfo
  {journal} {Phys. Rev. Lett.}\ }\textbf {\bibinfo {volume} {59}},\ \bibinfo
  {pages} {2044} (\bibinfo {year} {1987})}\BibitemShut {NoStop}%
\bibitem [{\citenamefont {G\"uhne}\ \emph {et~al.}(2007)\citenamefont
  {G\"uhne}, \citenamefont {Lu}, \citenamefont {Gao},\ and\ \citenamefont
  {Pan}}]{Otfried07}%
  \BibitemOpen
  \bibfield  {author} {\bibinfo {author} {\bibfnamefont {O.}~\bibnamefont
  {G\"uhne}}, \bibinfo {author} {\bibfnamefont {C.-Y.}\ \bibnamefont {Lu}},
  \bibinfo {author} {\bibfnamefont {W.-B.}\ \bibnamefont {Gao}}, \ and\
  \bibinfo {author} {\bibfnamefont {J.-W.}\ \bibnamefont {Pan}},\ }\href@noop
  {} {\bibfield  {journal} {\bibinfo  {journal} {Phys. Rev. A}\ }\textbf
  {\bibinfo {volume} {76}},\ \bibinfo {pages} {030305} (\bibinfo {year}
  {2007})}\BibitemShut {NoStop}%
\bibitem [{\citenamefont {T\'oth}\ and\ \citenamefont
  {G\"uhne}(2005)}]{Toth05}%
  \BibitemOpen
  \bibfield  {author} {\bibinfo {author} {\bibfnamefont {G.}~\bibnamefont
  {T\'oth}}\ and\ \bibinfo {author} {\bibfnamefont {O.}~\bibnamefont
  {G\"uhne}},\ }\href@noop {} {\bibfield  {journal} {\bibinfo  {journal} {Phys.
  Rev. Lett.}\ }\textbf {\bibinfo {volume} {94}},\ \bibinfo {pages} {060501}
  (\bibinfo {year} {2005})}\BibitemShut {NoStop}%
\bibitem [{\citenamefont {Zhang}\ \emph {et~al.}(2011)\citenamefont {Zhang},
  \citenamefont {Glancy},\ and\ \citenamefont {Knill}}]{Zhang11}%
  \BibitemOpen
  \bibfield  {author} {\bibinfo {author} {\bibfnamefont {Y.}~\bibnamefont
  {Zhang}}, \bibinfo {author} {\bibfnamefont {S.}~\bibnamefont {Glancy}}, \
  and\ \bibinfo {author} {\bibfnamefont {E.}~\bibnamefont {Knill}},\ }\href
  {\doibase 10.1103/PhysRevA.84.062118} {\bibfield  {journal} {\bibinfo
  {journal} {Phys. Rev. A}\ }\textbf {\bibinfo {volume} {84}},\ \bibinfo
  {pages} {062118} (\bibinfo {year} {2011})}\BibitemShut {NoStop}%
\bibitem [{\citenamefont {Laflamme}\ \emph {et~al.}(1996)\citenamefont
  {Laflamme}, \citenamefont {Miquel}, \citenamefont {Paz},\ and\ \citenamefont
  {Zurek}}]{Laflamme96}%
  \BibitemOpen
  \bibfield  {author} {\bibinfo {author} {\bibfnamefont {R.}~\bibnamefont
  {Laflamme}}, \bibinfo {author} {\bibfnamefont {C.}~\bibnamefont {Miquel}},
  \bibinfo {author} {\bibfnamefont {J.~P.}\ \bibnamefont {Paz}}, \ and\
  \bibinfo {author} {\bibfnamefont {W.~H.}\ \bibnamefont {Zurek}},\ }\href@noop
  {} {\bibfield  {journal} {\bibinfo  {journal} {Phys. Rev. Lett.}\ }\textbf
  {\bibinfo {volume} {77}},\ \bibinfo {pages} {198} (\bibinfo {year}
  {1996})}\BibitemShut {NoStop}%
\bibitem [{\citenamefont {Wang}\ \emph {et~al.}(2016)\citenamefont {Wang},
  \citenamefont {Chen}, \citenamefont {Li}, \citenamefont {Huang},
  \citenamefont {Liu}, \citenamefont {Chen}, \citenamefont {Luo}, \citenamefont
  {Su}, \citenamefont {Wu}, \citenamefont {Li}, \citenamefont {Lu},
  \citenamefont {Hu}, \citenamefont {Jiang}, \citenamefont {Peng},
  \citenamefont {Li}, \citenamefont {Liu}, \citenamefont {Chen}, \citenamefont
  {Lu},\ and\ \citenamefont {Pan}}]{Wang16arxiv}%
  \BibitemOpen
  \bibfield  {author} {\bibinfo {author} {\bibfnamefont {X.-L.}\ \bibnamefont
  {Wang}}, \bibinfo {author} {\bibfnamefont {L.-K.}\ \bibnamefont {Chen}},
  \bibinfo {author} {\bibfnamefont {W.}~\bibnamefont {Li}}, \bibinfo {author}
  {\bibfnamefont {H.-L.}\ \bibnamefont {Huang}}, \bibinfo {author}
  {\bibfnamefont {C.}~\bibnamefont {Liu}}, \bibinfo {author} {\bibfnamefont
  {C.}~\bibnamefont {Chen}}, \bibinfo {author} {\bibfnamefont {Y.-H.}\
  \bibnamefont {Luo}}, \bibinfo {author} {\bibfnamefont {Z.-E.}\ \bibnamefont
  {Su}}, \bibinfo {author} {\bibfnamefont {D.}~\bibnamefont {Wu}}, \bibinfo
  {author} {\bibfnamefont {Z.-D.}\ \bibnamefont {Li}}, \bibinfo {author}
  {\bibfnamefont {H.}~\bibnamefont {Lu}}, \bibinfo {author} {\bibfnamefont
  {Y.}~\bibnamefont {Hu}}, \bibinfo {author} {\bibfnamefont {X.}~\bibnamefont
  {Jiang}}, \bibinfo {author} {\bibfnamefont {C.-Z.}\ \bibnamefont {Peng}},
  \bibinfo {author} {\bibfnamefont {L.}~\bibnamefont {Li}}, \bibinfo {author}
  {\bibfnamefont {N.-L.}\ \bibnamefont {Liu}}, \bibinfo {author} {\bibfnamefont
  {Y.-A.}\ \bibnamefont {Chen}}, \bibinfo {author} {\bibfnamefont {C.-Y.}\
  \bibnamefont {Lu}}, \ and\ \bibinfo {author} {\bibfnamefont {J.-W.}\
  \bibnamefont {Pan}},\ }\href@noop {} {\bibfield  {journal} {\bibinfo
  {journal} {arXiv:1605.08547}\ } (\bibinfo {year} {2016})}\BibitemShut
  {NoStop}%
\bibitem [{\citenamefont {Takeuchi}(2001)}]{Takeuchi01}%
  \BibitemOpen
  \bibfield  {author} {\bibinfo {author} {\bibfnamefont {S.}~\bibnamefont
  {Takeuchi}},\ }\href@noop {} {\bibfield  {journal} {\bibinfo  {journal} {Opt.
  Lett.}\ }\textbf {\bibinfo {volume} {26}},\ \bibinfo {pages} {843} (\bibinfo
  {year} {2001})}\BibitemShut {NoStop}%
\bibitem [{not({\natexlab{g}})}]{note5}%
  \BibitemOpen
  \href@noop {} {\  ({\natexlab{g}})},\ \bibinfo {note} {the maximal collinear
  $d_{\text{eff}}^{\text{II}}$ is calculated to be equal to 2.02 pm/V in Ref
  .\cite{halevy11}.}\BibitemShut {Stop}%
\bibitem [{not({\natexlab{h}})}]{note6}%
  \BibitemOpen
  \href@noop {} {\  ({\natexlab{h}})},\ \bibinfo {note} {parts of similar
  results can been found in Ref. \cite{halevy11}}\BibitemShut {NoStop}%
\bibitem [{\citenamefont {Pinelis}(2006)}]{Pinelis06}%
  \BibitemOpen
  \bibfield  {author} {\bibinfo {author} {\bibfnamefont {I.}~\bibnamefont
  {Pinelis}},\ }\href@noop {} {\bibfield  {journal} {\bibinfo  {journal}
  {Electron. J. Probab.}\ }\textbf {\bibinfo {volume} {11}},\ \bibinfo {pages}
  {1049} (\bibinfo {year} {2006})}\BibitemShut {NoStop}%
\end{thebibliography}
\end{document}